\documentclass[preprint,12pt]{aastex}
\usepackage{graphicx}
\usepackage{xspace}

\newcommand{\kms}{~km~s$^{-1}$\xspace}
\newcommand{\mjy}{~mJy~beam$^{-1}$\xspace}
\newcommand{\jy}{~Jy~beam$^{-1}$\xspace}
\newcommand{\msun}{~$M_{\sun}$\xspace}

\newcommand{\lambdaone}{$\lambda = 1.3$\xspace}
\newcommand{\lambdathree}{$\lambda = 3$\xspace}
\newcommand{\csto}{C$^{17}$O\xspace}
\newcommand{\jto}{\mbox{J=2--1}\xspace}
\newcommand{\joz}{\mbox{J=1--0}\xspace}
\newcommand{\ethanol}{CH$_3$OH\xspace}
\newcommand{\water}{H$_2$O\xspace}
\newcommand{\nthp}{N$_2$H$^+$\xspace}
\newcommand{\tco}{$^{13}$CO\xspace}
\newcommand{\ceo}{C$^{18}$O\xspace}
\newcommand{\sot}{SO$_2$\xspace}
\newcommand{\spitzer}{{\it Spitzer}\xspace}
\newcommand{\um}{$\mu$m\xspace}

\begin{document}

\title{Kinematics of the Envelope and Two Bipolar Jets in the Class 0 Protostellar System L1157}

\author{Woojin Kwon\altaffilmark{1,2,3},
Manuel Fern{\'a}ndez-L{\'opez}\altaffilmark{2,4},
Ian W. Stephens\altaffilmark{2,5}, and
Leslie W. Looney\altaffilmark{2}}
\email{wkwon@kasi.re.kr}
\altaffiltext{1}{SRON Netherlands Institute for Space Research,
Landleven 12, 9747 AD Groningen, The Netherlands}
\altaffiltext{2}{Astronomy Department, University of Illinois, 1002
West Green Street, Urbana, IL 61801}
\altaffiltext{3}{Korea Astronomy and Space Science Institute, 776 Daedeok-daero, Yuseong-gu, Daejeon 34055, Republic of Korea}
\altaffiltext{4}{Instituto Argentino de Radioastronom{\'i}a, CCT-La Plata (CONICET), C.C.5, 1894, Villa Elisa, Argentina}
\altaffiltext{5}{Institute for Astrophysical Research, Boston University, Boston, MA 02215, USA}

\begin{abstract}

A massive envelope and a strong bipolar outflow are the two main structures
characterizing the youngest protostellar systems.  In order to understand
the physical properties of a bipolar outflow and the relationship with
those of the envelope, we obtained a mosaic map covering the whole bipolar outflow of the
youngest protostellar system L1157 with about $5''$ angular resolution
in CO \jto\ using the Combined Array for Research in Millimeter-wave Astronomy.  
By utilizing these observations of the whole bipolar outflow, we estimate its physical properties and show that they are consistent with multiple jets.  We also constrain a preferred precession direction. In addition, we observed the central envelope structure
with $2''$ resolution in the \lambdaone and 3
mm continua and various molecular lines: \csto, \ceo, \tco,
CS, CN, \nthp, \ethanol, \water, SO, and \sot.  All the CO isotopes
and CS, CN, and \nthp\ have been detected and imaged.  We marginally detected
the features that can be interpreted as a rotating inner envelope in \csto and \ceo and 
as an infalling outer envelope in \nthp. We also estimated the envelope and central protostellar masses and found that the dust opacity spectral index changes with radius.  

\end{abstract}

\keywords{
circumstellar matter ---
stars: formation ---
stars: individual (L1157) ---
stars: pre-main sequence ---
techniques: interferometric
}

\section{Introduction}  \label{sec:intro}

Bipolar outflows are the most energetic and remarkable phenomenon
of star formation.  Particularly, the youngest protostellar systems,
the so-called Class 0 young stellar objects (YSOs), are characterized
by a well-developed bipolar outflow with a massive envelope \citep{1993ApJ...406..122A}.  At
the earliest stage of star formation, material strongly accretes
onto the central protostar from the envelope, presumably through an accreting
disk. Meanwhile, a bipolar
outflow is launched and helps in removing angular momentum of the
accreting material.  Therefore, investigating the two main structures
(envelopes and bipolar outflows) together is
crucial for understanding the early stages of star formation and
evolution.  Note that what we call a bipolar outflow in this paper is a jet ejecting material rather than a slow molecular outflow mainly consisting of
the interacting features between the ejected jet and the ambient gas like a bow shock \citep[e.g.,][]{2014arXiv1402.3553F}. Because of this, the terms outflow and jet are used interchangeably in this paper.

Although much is known about bipolar outflows, such as their general physical and chemical properties, impacts on the environments, and
evolutionary changes \citep[e.g.,][]{2007prpl.conf..245A},
there remains many unanswered, fundamental questions:
What is the launching mechanism? Is it rotating? Is it
precessing and if so, what causes the precession?  Theoretical
studies agree that bipolar outflows are launched magnetocentrifugally
\citep{Blandford:1982uq,Ferreira:2006vn,Ferreira:1997kx,Cerqueira:2006ys},
but the launching mechanism and the launching regions are not yet clear \citep[c.f., X-wind
and disk-wind models,][]{Shu:1994dq,Konigl:2000bh}.  Bipolar outflows are expected to be rotating due to magnetocentrifugal launching and angular momentum conservation for the
accreting material.
Indeed, a small number of previous observational studies toward
Class 0 YSOs have reported a velocity gradient across bipolar outflow
widths, which can be interpreted as rotation
\citep{Pety:2006kl,Launhardt:2009fu,Zapata:2010dz,Choi:2011ly}.
However, different interpretations are possible such as multiple jet
events \citep{Soker:2013zr}.  In addition, bipolar outflows often
show a wiggling feature, which can be interpreted as precession.
However, with the exception of close binary systems, the precession mechanism has
not fully been understood \citep[e.g.,][]{Teixeira:2008fk}.

L1157-mm is one of the archetypical outflow sources. It is embedded in an
isolated star forming cloud located in Cepheus at a distance of 250 pc
\citep[][and references therein]{2007ApJ...670L.131L}.
With a bipolar outflow oriented nearly on the plane of the sky, L1157 is
one of the most extensively investigated Class 0 YSOs over
a wide range of wavelengths on various topics.  \citet{2007ApJ...670L.131L}
detected a flattened envelope structure in silhouette using the IRAC
data of the {\it Spitzer Space Telescope} (hereafter \spitzer).  \citet{2009ApJ...696..841K}
found that grains have significantly grown already at this earliest
stage, and the density distribution has a power-law index of $\sim
$1.8 toward the L1157 envelope.  Further studies at higher angular
resolution reported consistent results \citep{2012ApJ...756..168C}.
Recently, dust polarization has also been imaged by the Combined Array
for Research in Millimeter-wave Astronomy (CARMA)
\citep{2013ApJ...769L..15S}, which nicely shows a clear hourglass
morphology centered along the bipolar outflow.

The well-collimated bipolar outflow of L1157 was found over two
decades ago \citep{1992ApJ...392L..83U}.  Particularly the southern
blue-shifted lobe has been recognized to be heated by shocks,
and extensive molecular line studies have followed.
For example, time dependent shock chemistry was studied with a single-dish millimeter telescope \citep[e.g.,][]{1997ApJ...487L..93B,Bachiller:2001ju}.
Interferometric millimeter observations at $\sim3''$ resolution of CO and SiO 
in the southern lobe have suggested multiple bow shocks \citep[e.g.,][]{Gueth:1996ff,1998A&A...333..287G}.
In addition \citet{2010A&A...518L.120N} found that
in the shocked region, 20\% of all cooling is due to water molecules,
and \citet{2013A&A...557A..22S} distinguished
warm and hot components in the shocked regions
using the far-infrared spectroscopic imaging data of
the {\it Herschel Space Observatory}.

In this paper, we study the bipolar outflow and the envelope structure
of L1157 using CARMA. We report various
molecular line observations as well as continuum at \lambdaone\ and
3 mm. In particular, we obtained a mosaic map of CO \jto\ covering
the whole area of the L1157 bipolar outflow spanning over $5'$ with
$5''$ angular resolution.  In addition, based on a multi-jet model
fitting to the data cube, we find the physical parameters of the
bipolar outflow. We first describe
the details of the observations in Section \ref{sec:obs}. Then, the results
of the central envelope area are presented and discussed in Section
\ref{sec_envelope}.  In Section \ref{sec_outflow} and \ref{sec_discussion}, we show the
outflow mapping image, analyze the data, and discuss the physical
meanings. Finally, we enumerate the main conclusions in Section
\ref{sec:conclusions}.

\section{Observations and data reduction} \label{sec:obs}

We have obtained continuum and molecular line data toward L1157
using CARMA.  The data were taken between 2010 July and October.
An additional data set, originally taken as a CARMA summer school
project to carry out polarimetric observations toward L1157 in 2011
July \citep{2013ApJ...769L..15S}, has also been included.  The observations
are summarized in Table \ref{tab_obs}. For each molecular line
transition, Table \ref{tab_detect} shows the rest frequencies,
synthesized beam size, velocity resolution, and the root-mean square (RMS) noise per channel.

We have observed the entire bipolar outflow spanning over
$5'$ in CO \jto using 25 mosaic pointings.  When
designing the mosaic pattern, we used the \spitzer\ 8 \um\ image
\citep{2007ApJ...670L.131L}. All 25 mosaic points were observed
between each of the phase calibrator observations to
achieve a constant sensitivity over the entire field.  We have
employed the most compact CARMA E-array configuration in order to
maximize sensitivity for the largest scales possible, i.e., to
minimize the missing flux associated with the absence of zero spacing
visibilities for interferometric observations.  The configuration
provides an angular resolution of about $5''$ at \lambdaone\ mm and
is sensitive up to about $20\arcsec$ scales given the {\it uv}
coverage of CARMA \citep{2009ApJ...696..841K}.  Four bands
among 8 spectral bands available in CARMA were set to a wide bandwidth
of 500 MHz for continuum, and the other four bands were used to
observe the spectral lines CO, \csto, CN, \ethanol, and \water.  We
have configured the CARMA correlator with a bandwidth of 125 MHz
($\sim370$ \kms) and a velocity resolution of about 0.5 \kms\ at
\lambdaone\ mm for all the line observations.  The relatively broad
bandwidth guaranteed to cover all components of the bipolar outflow
with good velocity resolution.

We have also observed the central protostar of L1157 (L1157-mm) in the D-array 
configuration with a longer integration time at about $2''$ resolution.  We
employed the same correlator set-up as the bipolar outflow observations.

In addition, we have observed at  \lambdathree\ mm toward L1157-mm
in the C-array configuration to match the angular resolution of the
D-array data at \lambdaone\ mm. We used an 8 MHz band for \ceo\ and
\nthp\ (in the lower and upper sideband respectively) and four 31
MHz bands for the other molecular line transitions such as \tco,
\ethanol, SO, and \sot.  The last three bands were set to the wide
band of 500 MHz for observing continuum.

We also included the E-array data of the CARMA summer school project,
which had the primary goal of studying magnetic fields associated with
L1157.  The data consist of continuum, CO, and CS at \lambdaone\
mm.  The continuum data were used in the magnetic field studies by
\citet{2013ApJ...769L..15S} and the CO data were combined with our large CO data.  
We also report the results of the CS observations here.

Data have been calibrated following the general procedures using
the Multichannel Image Reconstruction, Image Analysis, and Display software
\citep[MIRIAD,][]{sault1995}.
The flux of the
gain calibrator 1927+739 was determined using a reliable primary flux calibrator
(Uranus or MWC349).  We obtained 1.7 Jy and 4.8 Jy for the 1927+739 fluxes
at \lambdaone\ and 3 mm, respectively, and used these values for all the data taken in 2010 since the data were taken over a small period of time.  On the other hand,
we used 1.2 Jy of the 1927+739 flux for the additional data set taken in 2011 which
was bootstrapped from its flux calibrator.
The absolute flux calibration uncertainties are about 15\% and 10\% at
\lambdaone and 3 mm, respectively \citep[e.g.,][]{2009ApJ...696..841K}.
When utilizing the target fluxes for estimating physical parameters in the paper,
we take into account only statistical uncertainties.
The natural weighting scheme was employed for image construction.  Note
that the molecular line data set from 2010 October 17 has been excluded for the final
combined images since the observations were taken in poor
weather conditions.

The final products of all the observational data are summarized in Table
\ref{tab_detect}.  As presented, the angular resolution toward the
central envelope is about $2''$ and for the bipolar outflow and
the summer school data (CS) about $5''$.  The velocity resolutions
and sensitivities of individual continuum and molecular lines are
also listed in the table.  We have not detected any structures in
\ethanol, \water, SO, and \sot\ for the achieved sensitivities.

\section{Central envelope region}  \label{sec_envelope}

To date, it is still nearly impossible to achieve high angular
resolution data with a good sensitivity for directly observing outflow launching
regions.  A reasonable approach, therefore, is to model the overall features
of the bipolar outflow and obtain an insight for the small scale features
such as launching regions.
In addition, the properties of the launching region may be related to
the physical properties of the envelope, which is the reservoir of the accreting material.
For example, the kinematics of the envelope are expected to match those of the bipolar outflow launching regions.  In this section we show and discuss the results of the envelope properties, and the observational and modeling results of the bipolar outflow are presented in the next section with an interpretation connecting the two structures.

\subsection{Continuum emission} \label{cont_emission}

\begin{figure}
\begin{center}
\vskip-5cm
\includegraphics[angle=0,scale=.60]{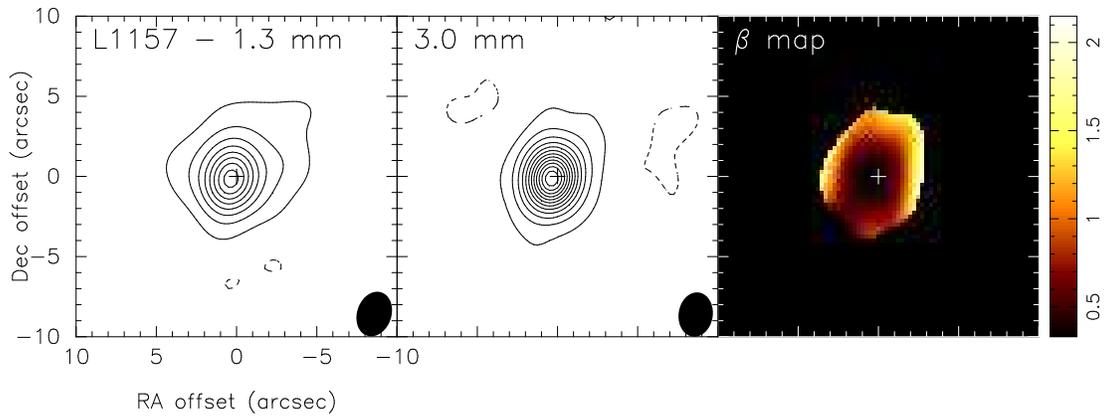}
\vskip-3cm
\caption{L1157 maps in continuum and the opacity spectral index $\beta$.
The contour levels are $\pm3$, 9, 15, 21, 27, 33, 39, 45, 51, 57, 63,
and 69 times 4.5 and 0.5 \mjy\ for \lambdaone and 3 mm, respectively.
The $\beta$ is in the range of 0.33 to 2.15.
The crosses indicate the phase center of our observations:
[R.A., dec.] (J2000) = [20:39:06.20, +68:02:15.90].
\label{fig_beta}}
\end{center}
\end{figure}

The \lambdaone and 3 mm continuum maps are presented in Figure
\ref{fig_beta}. In addition, the rightmost panel shows the dust
opacity spectral index ($\beta$) map with the assumption of optically thin emission: dust mass absorption coefficient $\kappa_\nu \propto \nu^\beta$.
As the Rayleigh-Jeans limit could be invalid at a low temperature even for millimeter wavelengths, we use the formula with the exponential terms of blackbody radiation:
\begin{equation}
\beta = \textrm{log}\Big(\frac{I_{\nu_1}}{I_{\nu_2}} ~ \frac{\textrm{exp}(h\nu_1/kT) - 1}{\textrm{exp}(h\nu_2/kT)-1}\Big)/\textrm{log}\Big(\frac{\nu_1}{\nu_2}\Big) - 3,
\label{eq_beta}
\end{equation}
where $I_\nu$ is an intensity at a frequency $\nu$, $h$ is the Planck constant,
$k$ is the Boltzmann constant, and $T$ is a temperature.
We adopted an envelope temperature of 20 K, which is reasonable for a Class 0 YSO envelope and consistent with radiative transfer modeling (private communication with H.-F. Chiang). 
At a temperature of 20 K, the $\beta$ value estimated using the Planck function is
$\sim0.2$ higher than that using the Rayleigh-Jeans approximation in the 1.3 and 3 mm
wavelengths.
The original data of the
\lambdathree mm continuum has a synthesized beam of
$2\farcs1\times1\farcs8$ (PA = $-74\degr$, a position angle measured counterclockwise from north) as listed in Table \ref{tab_detect}.  In order to compare with the \lambdaone mm continuum map and calculate the $\beta$ map
properly, we used a Gaussian-function tapering during map construction
and achieved a beam of $2\farcs8\times2\farcs1$ (PA = $-8\degr$),
which is nearly the same as the \lambdaone mm continuum map.  
The 1 mm and the tapered 3 mm continuum maps are shown 
in Figure \ref{fig_beta} with the synthesized beams marked at the bottom right.
The continuum peak measured by a 2D Gaussian function fitting is offset from the phase center to the southeast by $0.35\arcsec$ and is located at
R.A. (J2000) = $20^h39^m06.26^s$ and Dec (J2000) = $+68\degr 02\arcmin 15\farcs77$, which is consistent with higher angular resolution observations \citep[e.g.,][]{2012ApJ...756..168C}.

The total continuum fluxes at \lambdaone and 3 mm in a central
$10\arcsec\times10\arcsec$ box about the phase center are 
$0.472\pm0.018$ and $0.058\pm0.002$ Jy, respectively.
Note that the absolute flux calibration uncertainties are not included.
These continuum
fluxes provide an estimate of circumstellar material mass (circumstellar
disk and envelope, but mainly envelope mass) assuming the optically
thin case \citep[e.g.,][]{2000ApJ...529..477L}: $M \approx F_\nu D^2 / \kappa_\nu B_\nu(T_d)$ \citep{1983QJRAS..24..267H}, where $F_\nu$ is the integrated flux density, $D$ is the distance to L1157 of 250 pc, $\kappa_\nu$ is the mass absorption coefficient, and $B_\nu(T_d)$ is the Planck function at dust temperature $T_d$. We assume $\kappa_\nu$ is
0.01 \mbox{cm$^2$ g$^{-1}$} at \lambdaone mm \citep{1994A&A...291..943O},
corresponding to ice mantle grains with a size distribution with a power of 3.5
\citep{1977ApJ...217..425M} and the typical gas-to-dust ratio of 100.
Using a representative dust temperature of 20 K as addressed above, the envelope mass is 
$M_{env} \approx 0.583\pm0.022M_\sun$,
which  is consistent with previous studies
of radiative transfer modeling using a similar mass absorption
coefficient \citep[e.g.,][]{2009ApJ...696..841K}.
Free-free emission of low-mass YSOs is minimal (at most a few percent) at millimeter wavelengths, and in the case of L1157
it is approximated at a negligible level of 0.5--0.7 mJy at $\lambda = 1$ to 3 mm \citep{2013ApJ...779...93T}.
Note that the mass estimate could be uncertain up to about a factor of two due to uncertain mass absorption coefficients. The mass estimate error is purely from the statistical error of the total fluxes: absolute flux calibration errors up to 15\% (Section \ref{sec:obs}) is not included.
In addition, we estimated the deconvolved sizes by Gaussian fitting:
$2\farcs15\times1\farcs98$ (PA = $-4\degr$)  and $1\farcs68\times1\farcs54$
(PA = $-46\degr$) at \lambdaone and 3 mm, respectively.  The relatively more
compact size detected at the longer millimeter wavelength suggests
that the larger grains are centralized.
The flux and size estimates at \lambdathree mm are made in the tapered map 
explained in the previous paragraph.

\begin{figure}
\begin{center}
\vskip-5cm
\includegraphics[angle=0,scale=.70]{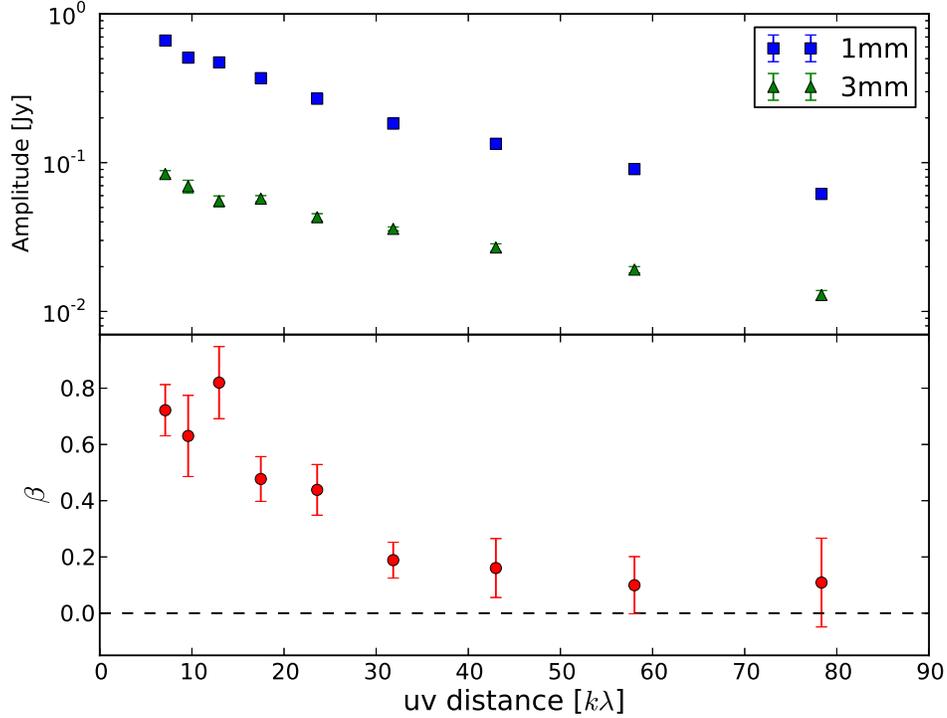}
\vskip-4cm
\caption{Amplitudes and $\beta$ along {\it uv} distances.  We obtained annulus
averages of both millimeter wavelength data in logarithmic bins 
and calculated $\beta$ in the optically thin case described in the text.  The error bars
indicate the $\beta$ ranges varying by the statistical standard
errors of annulus mean fluxes at \lambdaone and 3 mm.  Note that
the absolute flux calibration uncertainties (15\% and 10\% at
\lambdaone and 3 mm, respectively) can cause a systematic error up
to $\pm0.35$ in $\beta$ \citep{2009ApJ...696..841K}.
\label{fig_betauv}}
\end{center}
\end{figure}

The right panel of Figure \ref{fig_beta} shows how the derived dust opacity index, $\beta$, changes throughout the envelope of L1157 and is shown only at locations where both wavelength continua have a signal-to-noise higher than three.  
As shown in Figure \ref{fig_beta}, the central region of the envelope has a smaller
$\beta$ which is close to 0.3 (suggesting larger dust grains), and the boundary region
has a large value around 2 (similar to those estimated in the interstellar
medium, implying small sub-micron grains).  Note that $\beta$ is most sensitive
to the grain size \citep[e.g.,][]{2006ApJ...636.1114D}. This trend is understandable
in the sense that grains grow fast in a dense central region, and
the boundary has similar conditions to the interstellar medium, e.g.,
density and radiation field.  The mean
$\beta$, based on the total fluxes of the two frequencies, is 0.76.
In addition we present both wavelength data and $\beta$ with {\it uv} distances in Figure \ref{fig_betauv}, which also shows a variation of $\beta$.
We shifted the phase center to the continuum peak for the plot, and calculated $\beta$ by Equation \ref{eq_beta} assuming an optically thin case.
In fact, $\beta$ decreases with {\it uv} distance; the dense central
region has a smaller $\beta$ suggesting large grains.  Large optical depth can also cause changes in $\beta$; however, radiative transfer modeling could not explain such a large
variation of $\beta$ only with optical depth effects even in a more
massive envelope object \citep{2009ApJ...696..841K}.  Indeed,
although the uncertainties were large, data presented in
\citet{2009ApJ...696..841K} also show the same trend.  In contrast,
\citet{2012ApJ...756..168C} did not detect such a variation of
$\beta$ for the envelope of L1157.  We suspect that the discrepancy was caused by absolute
flux calibration uncertainties, especially when \citet{2012ApJ...756..168C} combined long baseline data taken in A and B configurations. 
In the common {\em uv} ranges, both data are in agreement within the absolute flux calibration uncertainties.

\subsection{Molecular lines}
\label{sec_mollines}

\begin{figure}
\begin{center}
\includegraphics[angle=0,scale=.6]{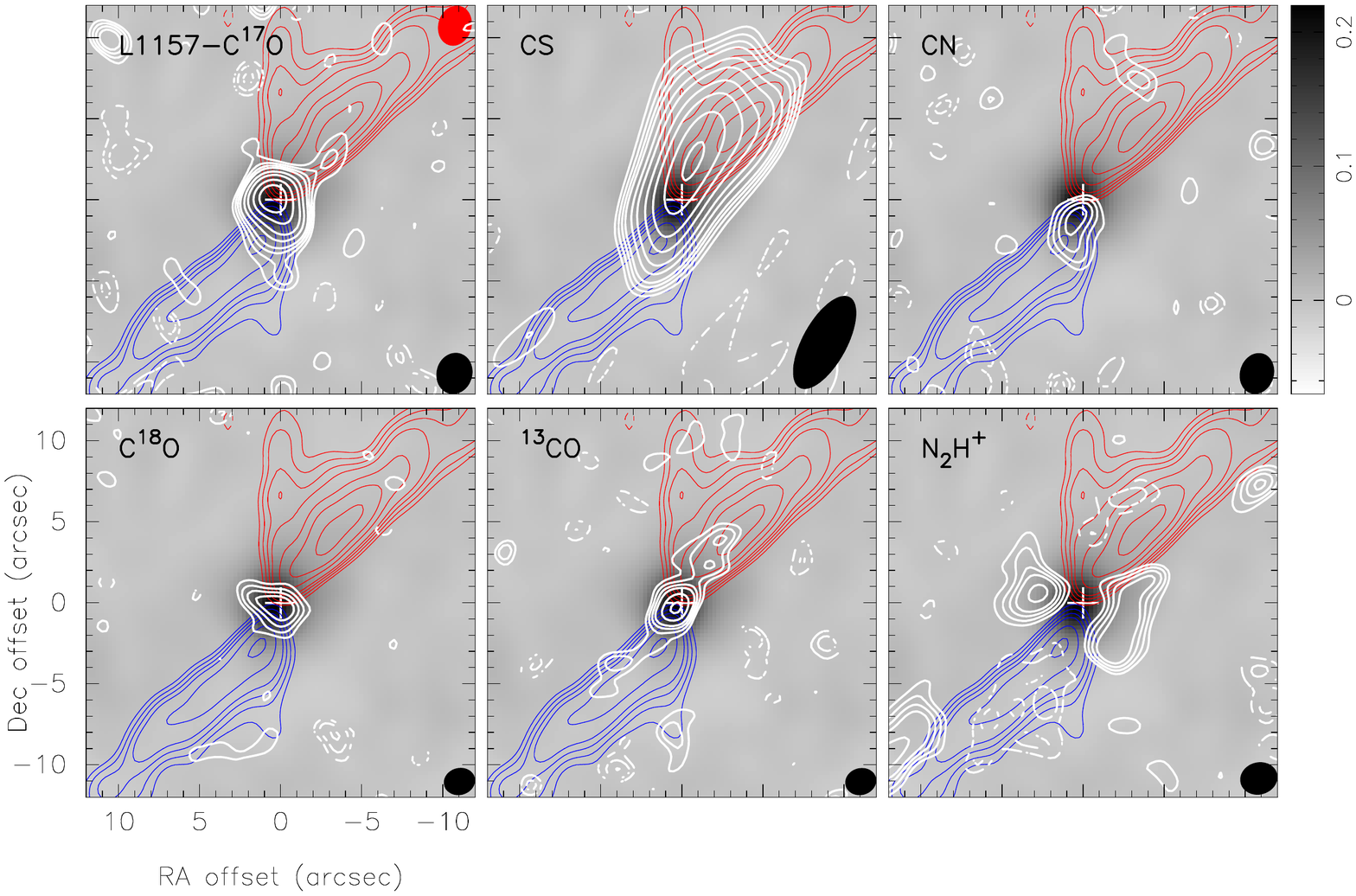}
\vskip-1cm
\caption{Integrated intensity maps of various molecular lines toward
the L1157 envelope.
The blue and red contours represent CO in velocity ranges of
--22 to 2 and 4 to 30 \kms, respectively.
All the contour levels are $\pm3$, 4, 5, 6, 8, 10, 13, 16, 20,
and 24 times $\sigma$: $\sigma = 2.0$ (CO),
0.080 (\csto), 0.30 (CS), 0.05 (CN), 0.045 (\ceo),
0.055 (\tco), and 0.050 (\nthp) \jy \kms.
The synthesized beams are also marked at the top-right for CO data
and the bottom-right for the others. The cross of each panel is the phase center.
The gray scales indicate \lambdaone\ mm continuum in units of \jy.
The velocity ranges integrated in individual lines are:
0.83 to 4.47 \kms for \csto, 1.38 to 4.17 \kms for CS, 1.25 to 3.25 \kms for CN,
1.35 to 3.95 \kms for \ceo, 0.35 to 5.22 \kms for \tco, and 0.59 to 3.80 \kms for \nthp.
\label{fig_m0map}}
\end{center}
\end{figure}

In this section, we present the results of various molecular line observations toward
the central envelope region with about 2$\arcsec$ resolution.
Figure \ref{fig_m0map} shows the integrated intensity maps of six molecular
lines overlaid with the CO \jto map\footnote{We indicate the rotational transitions only for CO and its isotopes. The spectral line details are found in Table \ref{tab_detect}.}.
The blue- and red-shifted components of the central region of the outflow are
shown in blue and red contours, respectively.
Since the bipolar outflow is nearly on the plane of the sky 
\citep[$i\approx81^\circ$,][]{Bachiller:2001ju}, the blue- and red-shifted
components are clearly distinguished on opposite sides.  In addition,
the integrated intensity peaks are not located in the middle of
both blue- and red-shifted lobes.  Instead, they trace two edges in a conic shape.

The three CO isotopes are differently distributed.  \csto \jto shows a
strong signal at the envelope and seems to trace the bipolar outflow
edges, presumably entrained gas. In contrast, \tco \joz is detected toward the envelope and the CO bipolar outflow peaks.  \ceo \joz does not trace
the bipolar outflow and instead shows the envelope elongated perpendicular
to the outflow.  
Interferometric observations suffer missing flux issues due to lack of short baselines. In the case of \tco \joz, the brightness temperature of single dish observations is 6 K at the line profile peak \citep{1997A&A...323..943G}, which corresponds to 
about 16 Jy per single-dish beam of $22''$. As the total flux of our data over the same region is about 2 Jy, the missing flux is approximately 87\%. As shown in Figure 5 and 6 of \citet{1997A&A...323..943G}, however, the large missing flux is mostly due to the large-scale low level emission. Missing flux is small toward the central peak and narrow jet features. For \ceo \joz and \csto \jto, whose features are likely more compact, the missing flux is expected to be even smaller.

\begin{figure}
\hskip-2cm \includegraphics[angle=0,scale=.7]{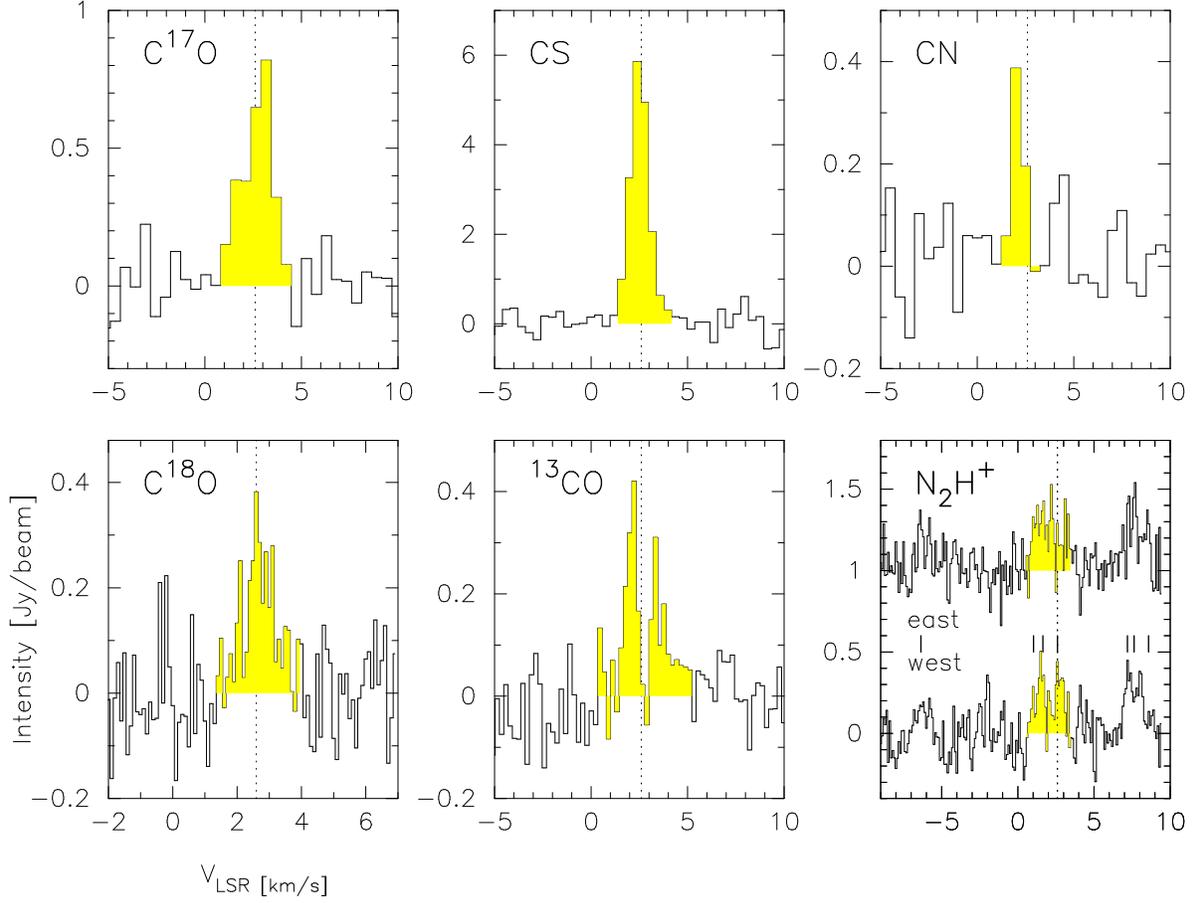}
\vskip-1cm
\caption{Line profiles averaged in a beam at the peak positions
of the integrated intensity maps presented in Figure \ref{fig_m0map}.  
The three CO isotope profiles have been taken at the center, as the line peak is consistent with the continuum peak.
For \nthp\ the integrated intensities of
a small area at the east and west blobs are shown:
a $3''\times3''$ box for the east and
a $2''\times3''$ box for the west blob.
The yellow shaded regions indicate the
velocity regions summed up for the integrated intensity maps.
For a better signal-to-noise, the \ceo\ and \nthp data have been
regridded in a velocity width of 0.1 \kms.
The dotted vertical lines indicate the $V_{LSR}$ of 2.6 \kms.
The short solid vertical lines in the middle of the \nthp panel mark the 7 hyperfine structures.
\label{fig_lineprofiles}}
\end{figure}

Differences among these three isotopes are also
found in the line profiles.  Figure \ref{fig_lineprofiles} shows the six line profiles at their
peak positions.  \tco \jto,
which arises from both the envelope and the bipolar outflow, has
a self-absorption feature (or extended features filtered by
the interferometric observations) and a broad wing component.  \csto \jto and
\ceo \joz have narrow line profiles, and a Gaussian profile fitting gives a full width half maximum (FWHM) of $1.82\pm0.19$ and $1.13\pm0.17$ \kms, respectively.
The hyperfine structures of \csto span over about 1.8 MHz: 224.7135334, 224.7141870, 
224.7147438, and 224.7153100 GHz \citep[the Cologne Database for Molecular Spectroscopy, ][]{Muller:2001ga}. However, the lowest frequency line at 224.7135334 GHz is negligible since its integrated intensity is lower than the others up to by an order of magnitude. Therefore, the effective frequency span of the hyperfine structures would be about 1.1 MHz, which corresponds to
about 1.47 \kms at the rest frequency. This explains the broad linewidth of \csto: $\sqrt{1.47^2 + 1.13^2} \approx 1.85$ \kms.
In Section \ref{sec_rotation}, we will also show that there exists a velocity gradient for \csto \jto and \ceo \joz about the envelope that may suggest rotation.

CS and CN also have clear detections.  However, their peaks are located
toward either the red- or blue-shifted outflow lobe.  The peak
of CS is north of the continuum center, which is toward the red-shifted
outflow lobe, and the peak of CN is south, at the blue-shifted lobe.
The line profile of CN has a FWHM of $0.73\pm0.18$ \kms and a velocity center
of $2.1\pm0.19$ \kms.  CN is the narrowest line presented in this paper and is found to be blue-shifted compared to the systemic velocity of 2.5 \kms.  The CS line profile is very
similar to \ceo in terms of the linewidth and the central velocity.
However the angular resolution of the CS data is relatively poor, so the peak
position is arguably consistent with the continuum peak.  
As the critical densities of
the two transitions we detected are similar
(1.1--1.4$\times10^6$ cm$^{-3}$), the discrepancy of CS
and CN distributions may indicate the differences in formation and/or
destruction.  For example, CS and CN both trace bipolar outflows, but CN
can be quickly destroyed, which results in a distribution
closer to the heating source \citep[e.g.,][]{1997ApJ...487L..93B}.
Note that single dish observations detected CS and CN over large
regions of the bipolar outflow \citep[e.g.,][]{Bachiller:2001ju}, but
our observations are limited only toward the center.

The distribution of \nthp clearly shows anti-correlation with
CO.  \nthp is found in the outer region of the envelope with a hole
at the center, and in the
southeastern region adjacent to the blue-shifted outflow lobe.

This anti-correlation is well understood by the fact that
\nthp is destroyed by CO \citep[e.g.,][]{2015ApJ...802....6S}.
As shown in Figure \ref{fig_lineprofiles}, 
the line profiles of \nthp around 
the east and west blobs marginally show multiple velocity components,
as expected for infall features.

\subsection{Infall of the outer envelope}
\label{sec_infall}

\citet{2010ApJ...709..470C} reported CARMA \nthp observations with an angular resolution of $7''$ and found that the \nthp traces well the flattened envelope structure \citet{2007ApJ...670L.131L} detected in silhouette by \spitzer.
They did not detect the central hole destroyed by CO at this angular
resolution, but they did detect a double-peaked feature toward the central region with velocities
around 2.7 and 3.1 \kms. They suggested that the two velocity components indicate
infall motion. In addition, a feature supporting a solid body rotation in a large scale of $\sim 20000$ AU was reported, which was argued as inherited from the large filamentary or flattened envelope kinematics.

\begin{figure}
\begin{center}
\vskip-5cm
\includegraphics[angle=0,scale=.70]{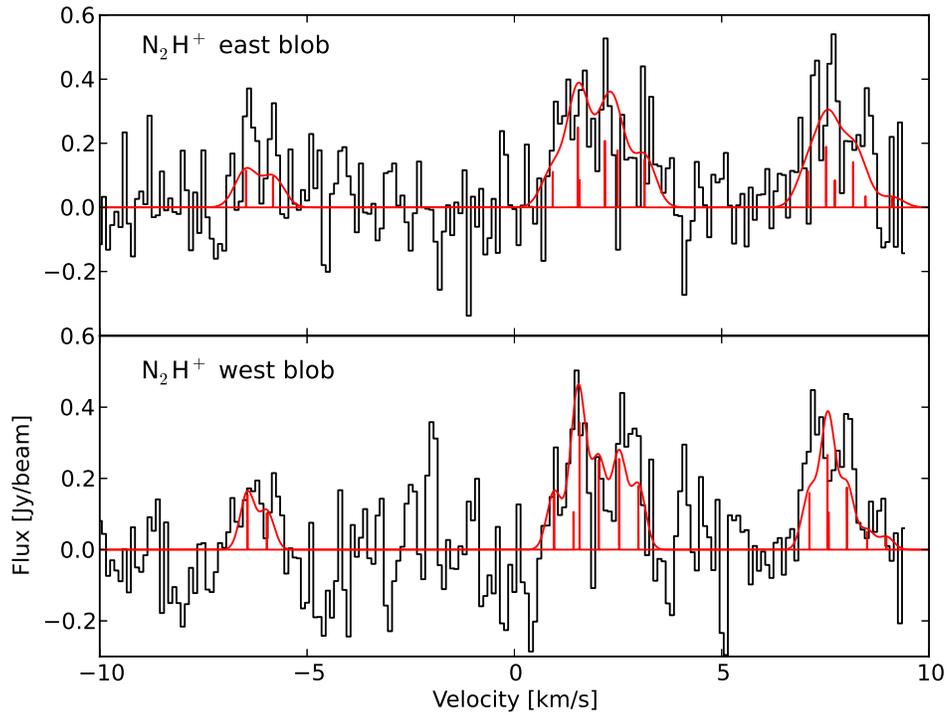}
\vskip-4cm
\caption{\nthp line fitting using two Gaussian profiles toward
individual east and west blobs. 
The leftmost component is the $JF_1F = 101 \rightarrow 012$ line.
\label{fig_nthpfit}}
\end{center}
\end{figure}

Our interferometric data with an angular resolution of $\sim 2''$ are not sensitive to 
the large scales of $\sim 20000$ AU, so the component interpreted as a solid body rotation by 
\citet{2010ApJ...709..470C} is filtered out. We convolved our \nthp channel maps with the identical synthesized beam of Chiang et al. and compared the $JF_1F = 101 \rightarrow 012$ hyperfine components (i.e., the more isolated hyperfine component in the \nthp spectra, located at a $V_{LSR}\sim -6.5$ \kms in Figure \ref{fig_lineprofiles}). We found that the line profile peak of our data is about 0.7 Jy beam$^{-1}$ while Chiang et al. obtained 0.9 Jy beam$^{-1}$, which implies that our observations filters out $\sim 22$\% of the \nthp flux observed by Chiang et al.
Despite the missing flux and poor sensitivity of our data, two velocity components are marginally detected toward
each of the east and west blobs. For each blob, individual
\nthp spectra were fit with two sets of Gaussian velocity components. The
7 hyperfine structures were fit simultaneously; their rest frequencies
and relative strengths are adopted from \citet{1995ApJ...455L..77C}
and  \citet{Womack:1992ek}, respectively.  Since the sensitivity
is marginal, we simply assumed optically thin emission and local thermal
equilibrium.  In addition, it is assumed that the two velocity
components have the same linewidth.  The fits results in velocity components of $2.6\pm0.04$
and $3.0\pm0.06$ \kms for the west blob and $2.5\pm0.07$
and $3.1\pm0.09$ \kms for the east (Figure \ref{fig_nthpfit}).
The FWHM is $0.4\pm0.06$ \kms in the west blob and $0.6\pm0.11$ \kms in the east blob.
The reported errors are statistical fitting uncertainties.  Note that the
spectra used for the fitting have a velocity resolution of 0.1 \kms.
\nthp is known to be optically thin for L1157, so the double peaks are unlikely
due to a self-absorption feature. Due to the low signal-to-noise at high spectral resolution, we also investigate whether a single velocity component can fit the data. The reduced $\chi^2$ value for the single velocity fit is almost identical to the two velocity fit.  However, the line profiles show obvious signs of two velocity components, particularly for the isolated, $JF_1F = 101 \rightarrow 012$ hyperfine component (Figure \ref{fig_nthpfit}), suggesting two velocity components may be meaningful.

These two velocities are consistent
with previous studies of lower angular resolutions discussing infall motion 
\citep[e.g.,][]{2010ApJ...709..470C}.
Since the two blobs present similar double velocity components
rather than a velocity gradient, the outer envelope ($\sim$1000 AU
scale) detected in our \nthp data is thought to be dominated by infall motion. 
The inner region of the \nthp hole
is nicely imaged with CO isotopes and consistent with a rotation feature (Section \ref{sec_rotation}).
The outside wall of the bipolar outflow cavity is also traced by
\nthp in more compact configurations
\citep{2010ApJ...709..470C,2012ApJ...748...16T}.
Hereafter, we will refer to the outer envelope as the region traced by \nthp\ from 500 to 1000 AU, and the inner envelope as the region traced by \ceo. Note that the dust continuum traces both these envelope scales. The outer envelope, as mapped by \citet{2010ApJ...709..470C}, extends to $\sim 20000$ AU scales, but again the observations in this paper cannot detect this large-scale structure due to spatial filtering of the interferometer.

\subsection{Rotation of the inner envelope}
\label{sec_rotation}

\begin{figure}
\begin{tabular}{@{\hskip-2cm}c@{\hskip-7cm}c}
\includegraphics[angle=0,scale=.50]{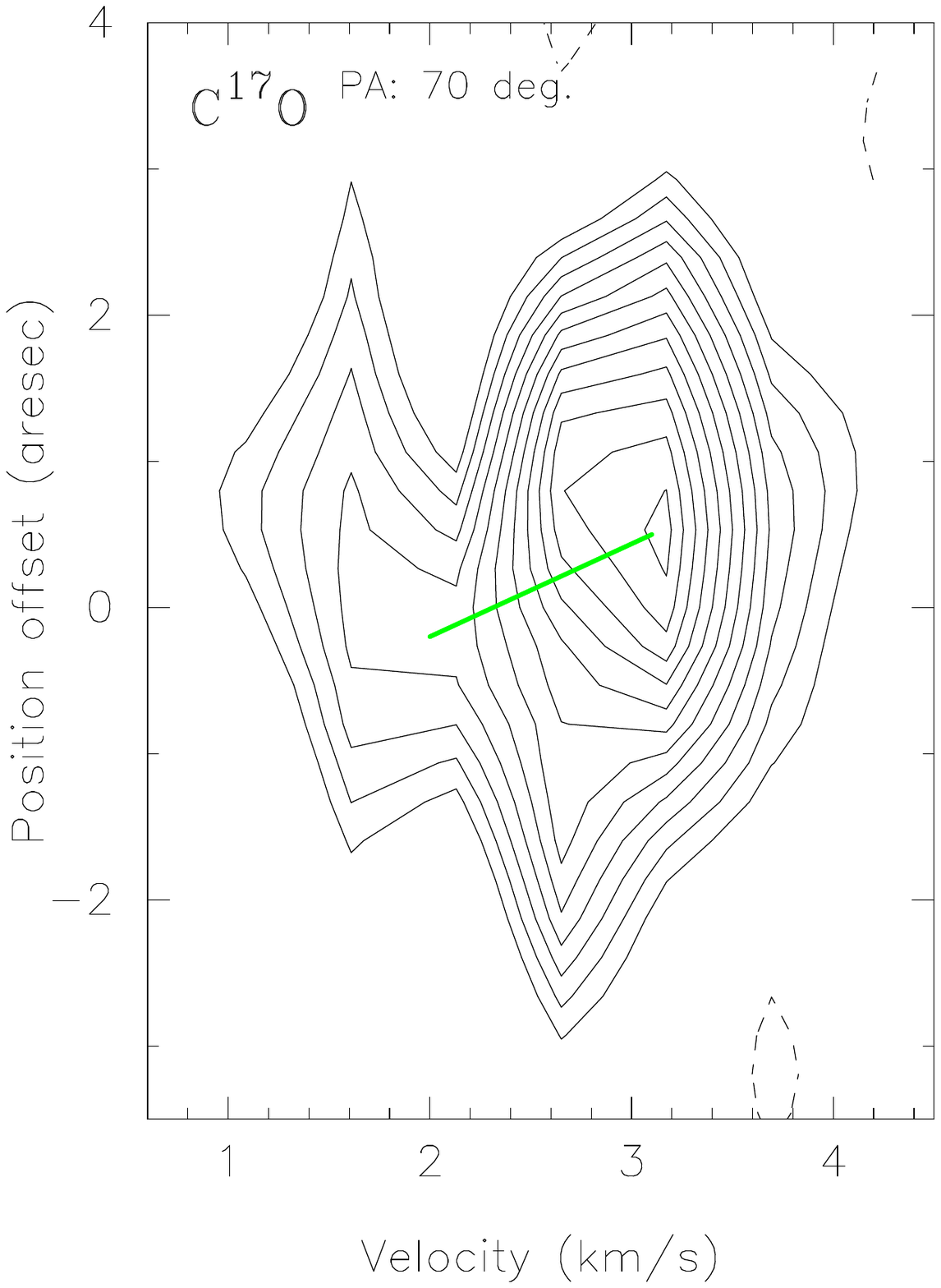}&
\includegraphics[angle=0,scale=.50]{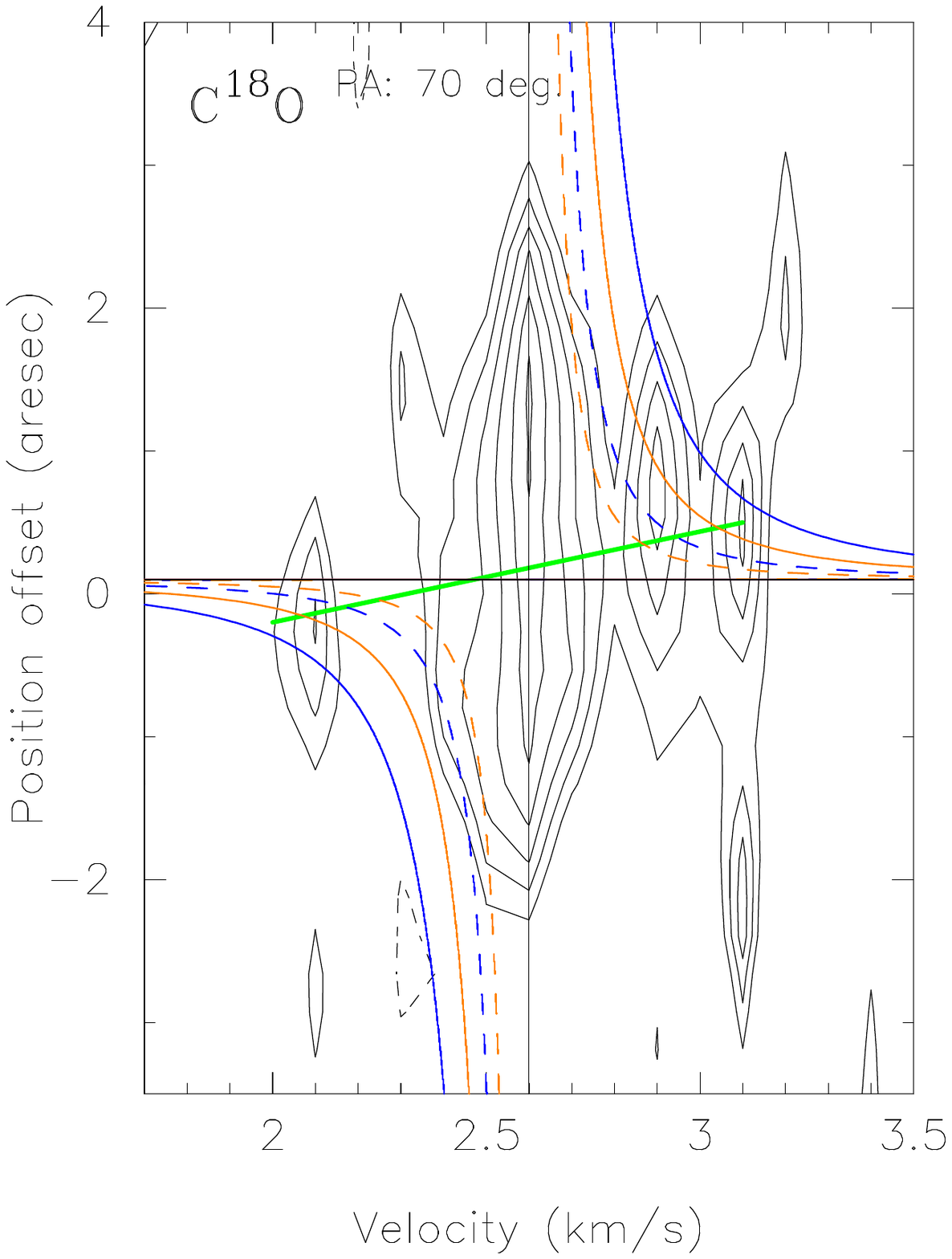} \\
\end{tabular}
\vskip-1cm
\caption{Position-velocity diagrams in \csto\ and \ceo\ 
along the line perpendicular to the bipolar outflow
(P.A. = $70\degr$) through the phase center.
The plus in position of the vertical axis is eastward.  The
lowest contours are $\pm0.18$ and $\pm0.15$ Jy beam$^{-1}$, and the
intervals are 0.06 and 0.05 Jy beam$^{-1}$ for \csto and \ceo,
respectively.
The vertical solid line
of the \ceo diagram marks $V_{LSR} = 2.6$ \kms and the horizontal
line indicates the location of the continuum peak, 
which is offset from the phase center by $0\farcs1$.
The green lines of the two panels connect the same
locations in position and velocity.  The blue and orange curves
indicate the Keplerian rotations around a central protostar of
$0.04~M_\sun$ and $0.02~M_\sun$ at a distance of 250 pc, respectively,
and the dashed lines are for the cases of a $60\degr$ inclination.
\label{fig_pv}}
\end{figure}

As shown in Figure \ref{fig_m0map}, CO isotopes well trace the envelope structure. In particular \ceo presents an elongated feature perpendicular to the bipolar outflow.
In order to investigate the kinematics of the inner envelope, we examined the position-velocity (PV) diagrams of \csto and \ceo cut through the phase center along two directions: 
the bipolar outflow (PA = $140\degr$) and the elongated envelope ($70\degr$).
The PV diagrams along the elongated envelope feature
show a velocity gradient, while the cut along the bipolar outflow does not.
The left and right panels of Figure \ref{fig_pv} present the \csto
and \ceo PV diagrams cut in the elongated envelope.
\csto shows double peaks in the PV diagram with a positional offset.
This feature is not due to the hyperfine structures of \csto. Unresolved hyperfine lines produce broader linewidths (Section \ref{sec_mollines}), and resolved hyperfine lines may show multiple peaks but not a velocity gradient. The velocity gradient corresponds to 1.1 \kms over $0\farcs7$ or 180 AU at the target distance of 250 pc.

The \ceo features are relatively complicated, but
the signals are mainly located in the top right and bottom
left in the diagram, which means that the eastern part of
the elongated envelope is redshifted and the western part
is blueshifted. This overall velocity gradient is in agreement with \csto. 
Note that the green lines in the two panels connect the same locations
in position and velocity.
The features shown in both \ceo and \csto could be a rotation, which is clockwise when looking down from the north. 
The 3.1 \kms component detected around an offset of $-2\arcsec$ is at the boundary of the outer envelope traced by \nthp, suggesting this could be infalling material.

The mass of the central objects can be estimated from the rotational
features of the molecular lines. Although our data are not high enough quality for clear evidence of Keplerian rotation, it is worthy to estimate the central object mass based on the interpretation.
In Figure \ref{fig_pv} we also
overlaid Keplerian rotation curves of a central protostar with a
mass of 0.04 and $0.02~M_\sun$ with blue and orange lines, respectively.
The mass of $0.04~M_\sun$ matches the overall outline of the \ceo
features in the PV diagram and the velocity gradients detected in
both \ceo and \csto.  
The mass of $0.02~M_\sun$ is
marked to show how much the curve changes with mass.  The bipolar
outflow is known to be nearly in the plane of the sky, which
presumably suggests that the rotation feature is close to edge-on.
Nevertheless, we show two dashed lines for the cases of $60\degr$ inclination from the plane of the sky.  The $60\degr$ inclination line is equivalent to half the
rotational velocities at a given radius.  In other words, if the
distance were indeed a half of 250 pc, the edge-on case is shown
by the dashed lines.  Similarly, the opposite case of a 500 pc
distance can simply be estimated in the plot.

For the Keplerian rotation curves, we did not take into account the
envelope mass within a given radius: $M(R) \approx
M_{env}(R/R_{out})^{3-p}$, where $M_{env}$ is the total envelope
mass (0.58 $M_\sun$, as derived in Section \ref{cont_emission}),
$R_{out}$ is the envelope size ($5''$), and $p$ is the power-law
index of a volume density distribution.  When assuming $p=1.8$
\citep{2009ApJ...696..841K}, the envelope mass inside $0\farcs35$
(the radius of the offset between the two peaks in the PV diagrams
connected by a green line in Figure \ref{fig_pv}) is 0.02 $M_\sun$.
Due to the cavity swept by the bipolar outflow and the sharply
increasing dust temperature close to the center, the mass inside is
likely smaller than this estimate.  This
implies that Keplerian rotation tends to appear inside around
$0\farcs35$ where the envelope mass portion can be neglected, while
a ``constant'' velocity might appear beyond $0\farcs55$ ($\sim140$
AU at a distance of 250 pc) where the envelope mass inside is
comparable to the protostellar mass ($0.04~M_\sun$): $v =
(GM(R)/R)^{1/2}$, i.e., $v \propto R^{2-p}$.  Again, on the scales of
1000 AU, infall motion is detected in \nthp.

Although it is not clear due to the limitation of our data in
sensitivity and resolution, Figure \ref{fig_pv} indicates that the
outer contours and the overall gradient in the green line can be
explained by Keplerian rotation around a $0.04~M_\sun$ protostar.
Note that we adopted the distance of 250 pc, which is most reasonable
\citep{2007ApJ...670L.131L}, but the mass estimate decreases or
increases linearly depending on the distance.  \citet{1997A&A...323..943G}
estimated the central protostellar mass as $0.2~M_\sun$ based on
the velocity gradient detected in \ceo, but they assumed the distance
of 480 pc and overestimated the interval across the 1 \kms velocity
gradient roughly as $\sim 3''$, which both cause an overestimate
of the protostellar mass.  
Since \citet{1997A&A...323..943G} interferometric observations have a comparable angular resolution to our data, missing flux might not cause a significant bias. It is not easy to compare the two interferometric data sets since they have different channel widths and a weighting. Their Figure 9 presents a channel map with uniform weighting and shows intensity peaks separated by $3''$, but the span only appears well in two adjacent channels (2.81 and 3.02 km/s), not blue- and red-shifted channels. On the other hand, their Figure 4 that uses a natural weighting shows a similar position span for the velocities we observed. Considering the uncertainties and differences of both data sets, we argue that our interpretation is not so problematic, and that
the two estimates are consistent.  Based on the dust continuum, we estimated
the envelope mass as $\sim0.58~M_\sun$ (Section \ref{cont_emission}), which gives an 
envelope to protostellar mass ratio of about 14.

\section{Bipolar outflow} \label{sec_outflow}

\subsection{Outflow activity}
\label{sec_outflowactivity}

\begin{figure}[h]
\epsscale{0.9}
\plotone{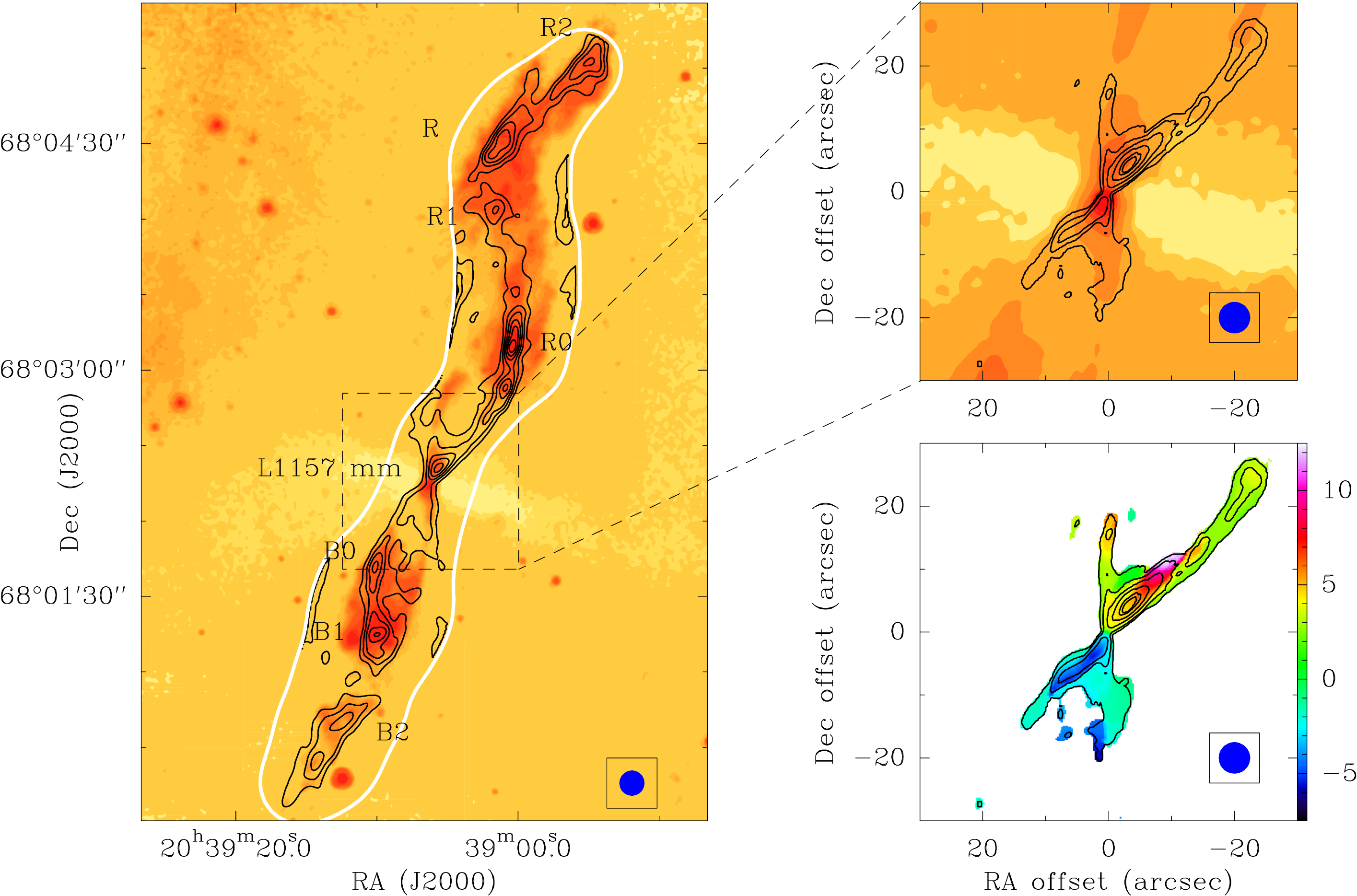}
\caption{{\bf Left:} CO  CARMA E-array configuration image
(black contours) on top of the 8~$\mu$m \spitzer image of the L1157
outflow system. The contours are 3, 8, 13, 18, 23, 28, 33 and 38 $\times$
3.0~\jy \kms (the rms noise level is $\sim$1.3\jy \kms). The
positions of knots R, R0, R1, R2, B0, B1, and B2 are marked, as well
as the central millimeter continuum source. The synthesized beam
is plotted in the bottom right corner. The white solid line marks the area covered by the half-power primary beams of the 25 pointing mosaic.
{\bf Top right:} Zoom-in of
the central part of the L1157 system. The contours show the combined
CO  CARMA E+D-array configuration image on top of the 8~$\mu$m
\spitzer image. Contours are 3, 15, 35, 55, 75 and 85 $\times$ 1.0\jy \kms,
the rms noise level. The synthesized beam is shown in the bottom
right corner. {\bf Bottom right:} Same as above but showing the
velocity centroid (moment 1) map of the CO  emission (color
scale) overlaid with the integrated intensity moment 0 image
(black contours). 
} 
\label{mom0} \end{figure}

The left panel of Figure \ref{mom0} shows the CO emission
integrated in velocity (moment 0) observed in the E-array configuration
(black contours) at a 5$\arcsec$ spatial resolution overlaid on
the 8~$\mu$m emission taken by the \spitzer\ telescope
\citep{2007ApJ...670L.131L}. The figure clearly shows overlap of
the 8 $\mu$m and CO emission.  Note that the \spitzer 8~$\mu$m
channel may be dominated by H$_2$ 0--0 S(5) and S(4) tracing shocked
gas \citep{2007ApJ...670L.131L}. The CO emission indicates a wiggling
bipolar morphology running northwest-southeast, which is a well-known
indication of precession that is also detected in several low energy
SiO transitions \citep[e.g.,][]{Zhang:2000ye}. Within about 30$\arcsec$
from the central protostar, the CO
emission shows a point symmetric X-shape morphology composed of
two narrow ridges of $\sim3\arcsec$ width. At further distances, the CO emission
consists of several bright knots along an elongated curved structure spanning in total $\sim300\arcsec$ (75000~AU). Most of the
bright knots have been extensively reported and studied in the past
(e.g., Bachiller et al. 2001), including B0, B1 and B2 in the
blue-shifted (southern) lobe and R, R0, R1 and R2 in the red-shifted (northern)
lobe as indicated in Figure \ref{mom0}.

The upper right panel of Figure \ref{mom0} is a zoom-in of the central
region, showing contours of the CO emission observed in CARMA D
configuration at $2\farcs5$ resolution. The color scale once
again shows the \spitzer\ 8 $\mu$m emission and the yellow tones
indicate the well-known extinction lane perpendicular to the outflow.
The observations show a clear X-shape pattern, with the strongest
emission in the northwest and southeast arms. The average FWHM of
these arms is about $3\farcs3$, which is about $2\farcs2$ (550~AU) in width
after deconvolving with the beam.  The lower right panel
of Figure \ref{mom0} shows the CO centroid velocity (moment 1) map
of the central region. The velocity in this map ranges from 
--7.5\kms to +12.5\kms 
(note that the velocity is indicated with respect to 
the cloud velocity $V_{LSR}=+2.6$\kms, section
\ref{sec_rotation}).  By inspecting the image, it is evident that
close to L1157-mm, the northwest-southeast arm of the X-shaped
structure has an average absolute velocity 2--3\kms higher than the
northeast-southwest arm. The arms of this X-shaped emission has
been interpreted as the limb-brightened cavity walls of the outflow
\citep[e.g.,][]{1997A&A...323..943G, 2013A&A...558A..94G}.  However,
our data show certain similarities between the northwest and southeast arms in morphology, symmetry, brightness,
and kinematics of the emission, suggesting components of one jet. The same occurs with the northeast-southwest arms, though fainter, suggesting these arms could be another jet.  In Section \ref{sec:jetsoutflows}
we discuss the possibility that the CO emission is associated with
multiple outflows or molecular jet ejections from the central
disk-protostar system.

Even though CARMA is filtering out the fainter emission between the two bright narrow ridges shown by the single-dish and interferometer combined CO \joz\ map of
\citet{Gueth:1996ff}, which interpreted the system as a limb-brightened cavity,
the possibility of two jets is not ruled out. \citet{Gueth:1996ff} reported a very high emission contrast between the ridges and the region enclosed by the ridges, which they could not completely model even using an exponential decay in the density profile of the putative cavity.
The compactness of ridges allows CARMA to detect them:
the largest angular scale of CARMA in this configuration goes up to $\sim20\arcsec$. The proposed multi-jet model does not need to recover all the extended flux to be tested and has the advantage of explaining several observed features that cannot be explained by the cavity model. As mentioned in \cite{Gueth:1996ff}, the cavity model has a difficulty in explaining (1) the edge-inner cavity emission contrast, (2) the morphology of the emission at high velocity showing only the cavity walls, and (3) the non-homogeneous emission of the inner cavity showing several arches and small filaments.

\begin{figure}[h]
\vskip-5cm
\plotone{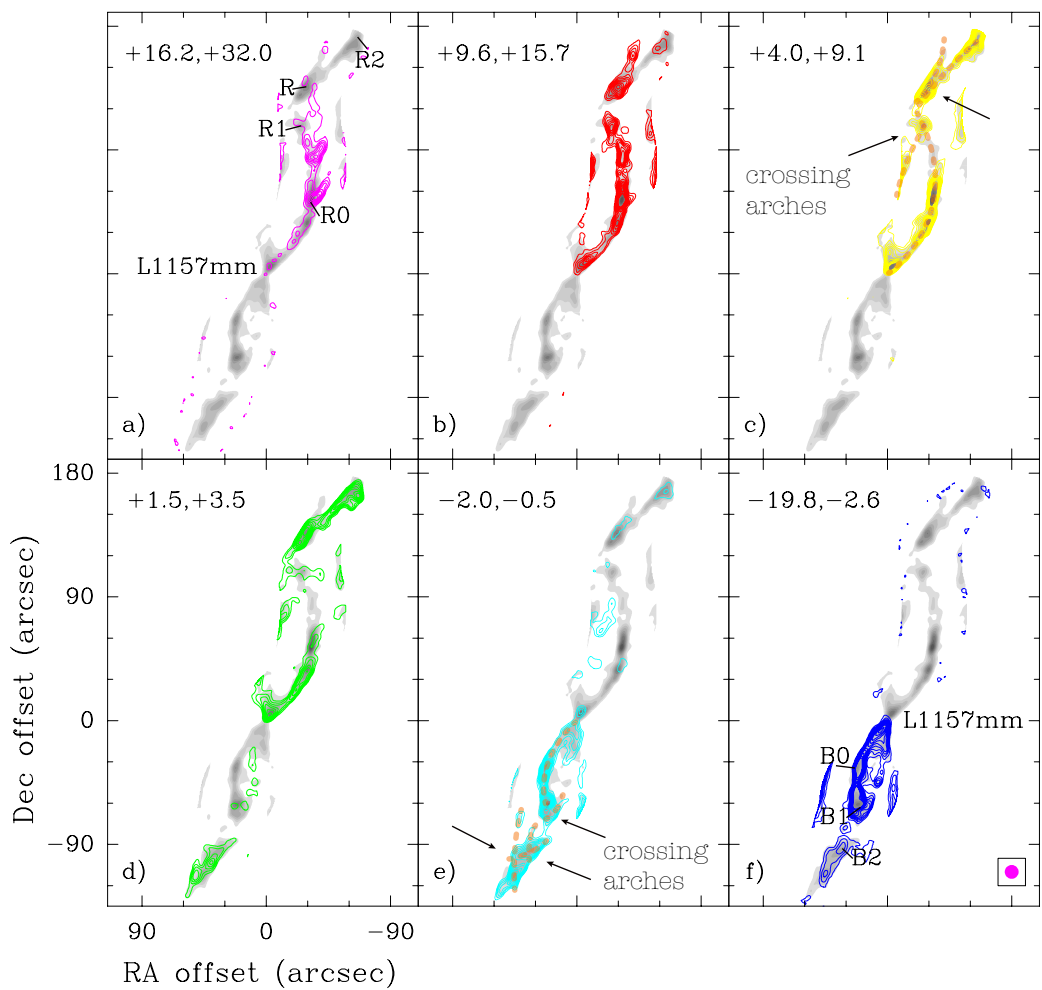}
\caption{CO  emission velocity map (contours) overlapped with
the integrated emission (grey scale). The contours are
3, 6, 9, 12, 15, 18, 21, 24, 27 and 30 $\times$ the rms noise level of each
channel (0.7, 0.4, 0.3, 0.3, 0.4 and 0.7\jy \kms, for a-f panels respectively). The synthesized beam is shown in the
bottom right panel and the velocity interval over which each channel
was integrated is indicated in each panel. In panels c) and e) the path of some arches of emission are marked with dashed orange lines.} \label{cube} \end{figure}

Figure \ref{cube} shows the CO emission integrated over 6
arbitrarily selected velocity intervals so that the overall morphology
and kinematics of the outflow are adequately depicted. The extended
emission at velocities close to the systemic velocity is filtered
by the interferometer and hence not shown. Although the rest of the
velocity channels are also affected by the presence of 
extended emission, the main issue is the dynamic range of the 
images since the narrow outflow ridges are much brighter than 
the emission from the putative inner cavity material 
\citep{Gueth:1996ff}. Nevertheless, we have checked that the sidelobes 
have little impact in the brightness distribution of the ridges themselves.
We also noticed in the CO moment 0 image, there are some artifacts at the edges of the images. In particular, the emission features west of R1 and R0 and west and east of B1 are sidelobes of the strongest outflow ridge. These edge problems only appear in the channels (Fig. \ref{cube})  where this ridge is stronger. On the contrary, the emission southeast of R1 is clearly not an artifact. One can see for instance, that its curvature is the opposite from the main northwest ridge passing through R0.

The redshifted high velocity CO emission (+16.2,~+32.0\kms ; Fig. \ref{cube} a)
traces part of the northwest outflow lobe with some finger-like
features protruding northwest from the curved molecular jet path. At
medium velocity (+9.6,~+15.7\kms ; Fig. \ref{cube} b) the CO emission runs mainly
from the protostar position and extends northward to knot R along a western 
curved ridge. There is also emission along the eastern curved 
feature southeast of R1, although this feature was not well mapped due to the limited field of view of our observations. 
Some emission is also detected north of knot R. The redshifted low velocity
emission (+4.0,~+9.1\kms ; Fig. \ref{cube} c) is similar to that of the medium 
velocity emission. The emission runs outward from the protostar along
the knots R0, R1, R, and R2.  CO emission is also seen northeast of
L1157-mm and southeast of R1 which may indicate a different curved
jet path. This velocity channel shows emission at knots R and R2.
Also evident is that the curved features cross each other at the infrared knots R and R1. Close to the cloud velocity (+1.5,~+3.5\kms ; Fig. \ref{cube} d), the emission
associated with knots R0 and R1 becomes fainter, while that at knot B2 in the southern blue lobe becomes obvious.
For the blueshifted low velocity (--2.0,~--0.5\kms ; Fig. \ref{cube} e), knot B2 shows an extended morphology. Hints of arched structures apparently crossing each other are seen at B1 and B2. At both low and high blueshifted velocities, the
main southeast curved ridge is seen from L1157-mm and
along knots B0 and B1. 
In addition, there is a blueshifted emission component in the middle of the northern lobe at $\sim60\arcsec$ north of L1157--mm. At high blueshifted velocities (--19.8,~--2.6\kms ; Fig. \ref{cube} f), an arch joining L1157-mm with the western part of B1 forms the southwest curved ridge.

We do not include velocity dispersion (second moment) maps in this paper because the spectral profiles of the CO emission show extended emission wings that appear masked by the core emission of the line profiles. Pronounced wings (extending up to $\sim33$\kms) are commonly found along all the knots marked in Fig. \ref{mom0}, while the rest of the gas have linewidths between (1--3) \kms.

\begin{figure}[h]
\epsscale{0.9}
\vskip-3cm
\plotone{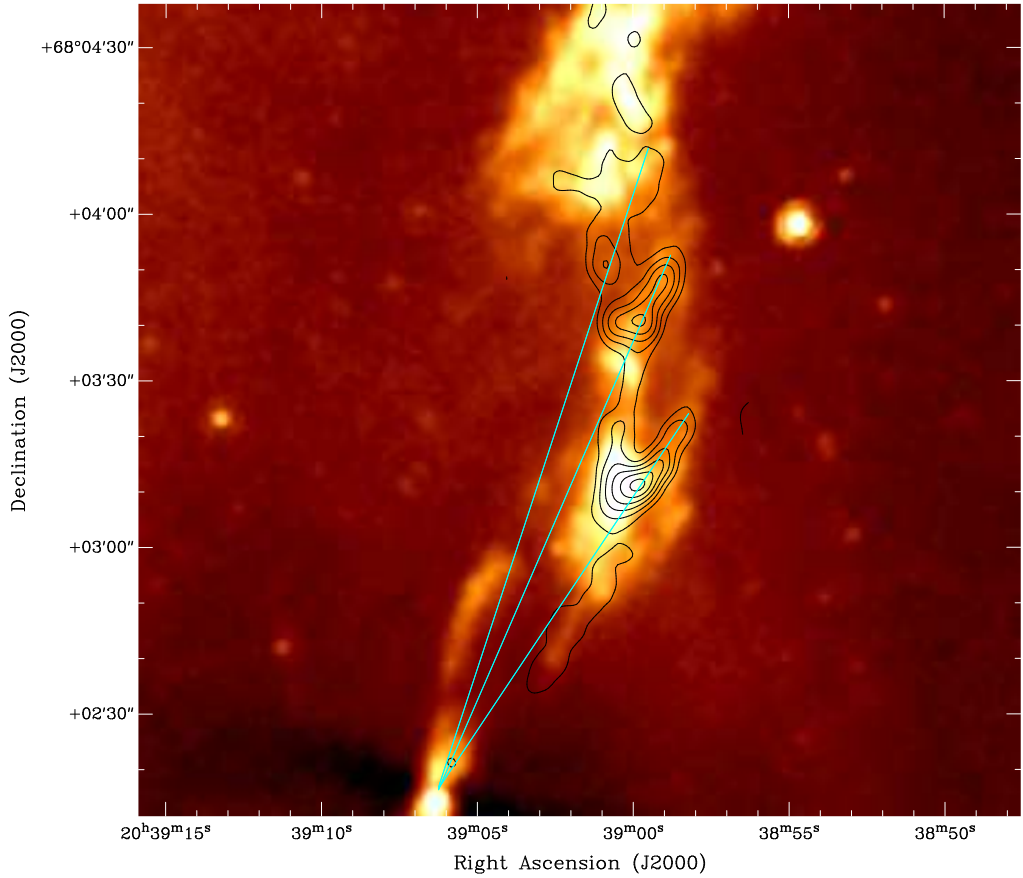}
\vskip-3cm
\caption{Zoom-in of the northern lobe of the L1157 outflow system.
The contours show the most redshifted CO  emission and the
color scale is the \spitzer 8~$\mu$m. The contours are
15, 30, 45, 60, 75, and 90\% of the peak, 15.4 \jy \kms.
The blue lines indicate the
ballistic trajectories of three possible molecular bullets ejected
at different position angles.} \label{bullets} \end{figure}

Figure \ref{bullets} shows a zoom-in of the north lobe of the
outflow. It shows the infrared 8~$\mu$m emission overlaid with
the CO redshifted high velocity emission between +32\kms and
+18\kms. This image shows two clear finger-like features at
140$\arcsec$ and 150$\arcsec$ from the L1157-mm position at a PA
of --60$\arcdeg$ and --50$\arcdeg$ respectively. There is also possibly
a third finger at 155$\arcsec$ away from L1157-mm at  PA $ = -40\arcdeg$.
The tip of these fingers spatially coincide with some of the 
mid-IR knots studied by \citet{2011Takami}.
All three features are marked in the figure with a straight line
joining each of them with the position of L1157-mm. Their linear
trajectory along with their high velocity suggests that they can be
molecular bullets ejected with different position angles.
Therefore, these fingers indicate a possible rotation and precession of the ejecting source.

\subsection{Two molecular jets in L1157} \label{sec:jetsoutflows}
\label{tjets}

Previous work based on molecular observations towards L1157
\citep{Gueth:1996ff,Gueth:1998lh,Bachiller:1997fu} have proposed the
existence of an episodic precessing jet which drags the environment
and excavates a wide cavity. The cavity walls are limb-brightened
and hence easily detected in molecular emission. However, the L1157
outflow shows a point reflection symmetry of the strongest clumps
at the northwest and southeast putative cavity walls. The symmetry
of the walls is seen via three physical parameters: shape, radial
velocity, and clump structure.  Thus, previous investigations
\citep{Zhang:2000ye,Bachiller:2001ju} indicated the possibility that
not only the CO emission but also the SiO emission (a well known
tracer of the shocked gas in molecular outflows and jets) in L1157
is tracing one narrow molecular precessing jet, coincident with the
northwest-southeast CO arm seen close to the protostar.
The idea of a precessing jet is reinforced by the finding of
high velocity shocks at different position angles in the north
lobe of the outflow (see Section \ref{sec_outflowactivity}).
\citet{Bachiller:2001ju} also speculated the existence of a milder
wide-angle wind surrounding this narrow precessing jet, which would
account for the weak CO molecular emission (undetected in SiO) to
the northeast of the red (northern) lobe and the southwest of the blue (southern) lobe. However, this
scenario can poorly account for this \lq\lq anomalous\rq\rq emission
since it is not detected at both sides of the outflow simultaneously
(i.e., there is no CO emission west of the narrow SiO molecular
jet in the north lobe and to the east in the south lobe). Moreover,
the velocity of a wide-open angle wind is expected to decrease with
the opening angle from the outflow axis \citep{Arce:2004qo}. 
However, such a characteristic is not found in our data.

On the basis of the new CARMA CO observations, we explore the possibility
of the ejection of multiple jets from the L1157 protostar. In
particular we fit a two precession jet model to the CO data. The idea of a
precessing jet needs a mechanism to eject material that follows
ballistic trajectories. Evidence for high velocity ballistic ejections
are clearly shown in Figure \ref{bullets}. In addition, there is
evidence suggesting that some of these bullets are cometary-shaped
and run parallel to each other in the southern lobe \citep{2011Takami}.
This matches well with the idea of multiple jets or ejections being present in
the outflow.

\begin{figure}[h]
\epsscale{0.5}
\plotone{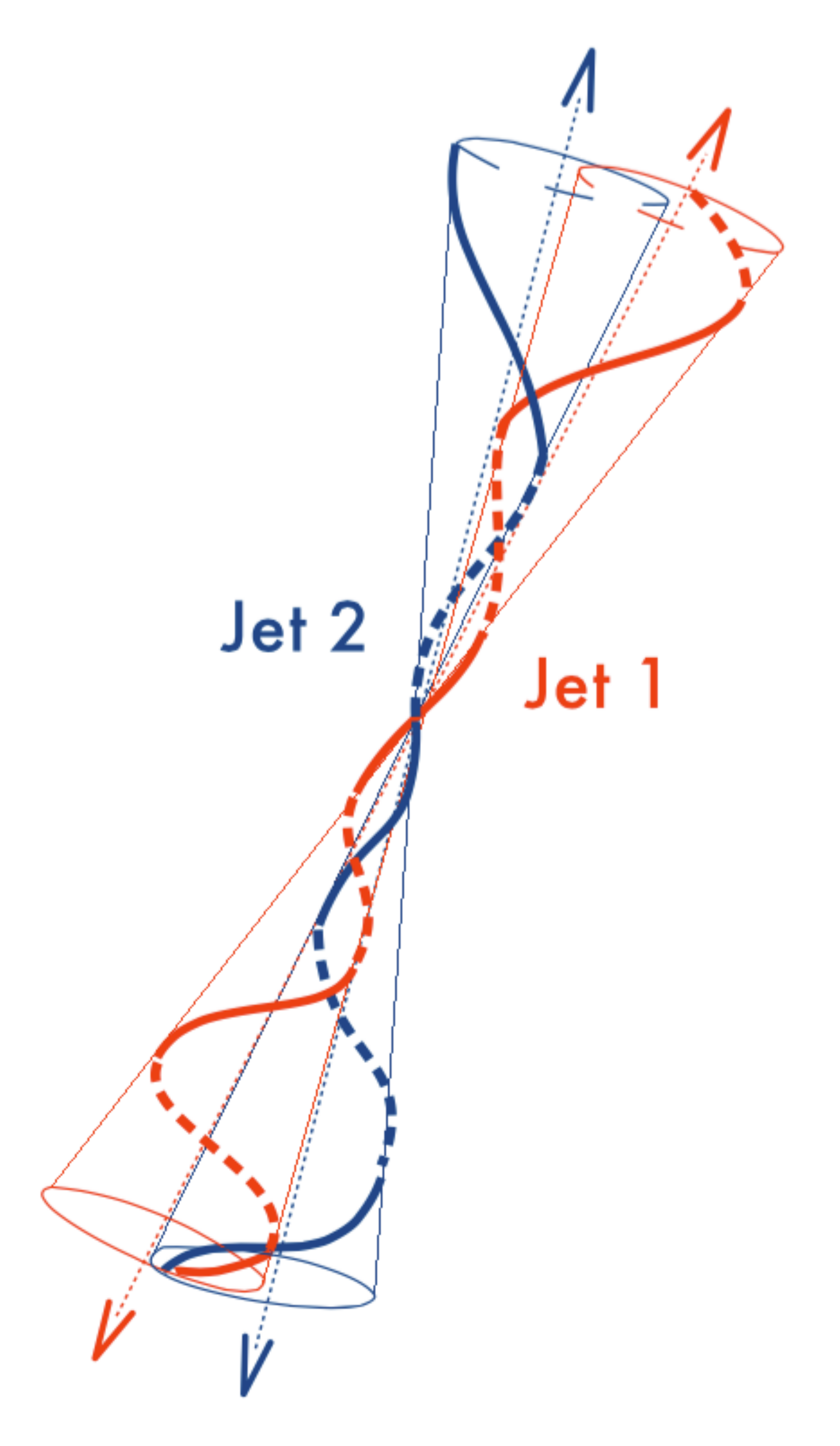}
\caption{Cartoon showing the new interpretation for the L1157 outflow system as
including multiple precessing molecular jets launched at different directions. 
Dashed lines show the path of the jets on the rear part of the cones, while solid lines show the path on the front part of the cones. The pointed arrows show the axis of symmetry of the two bipolar outflows.
} \label{sketch} \end{figure}

We suggest that the initially supposed limb-brightened cavity edges 
\citep[e.g., ][]{Gueth:1996ff,Gueth:1998lh}
can possibly be the path of two molecular jets (see Figure 
\ref{sketch} for a sketch of this new interpretation). 
The CO CARMA images
(Figures \ref{mom0} and \ref{cube}) show a clear wiggling path
running northwest-southeast close to the central protostar.  It
coincides well with the previously identified SiO molecular jet,
which we call {\it Jet 1}.  The CO  emission shows also another
S-shaped feature running northeast-southwest close to the protostar
which we call {\it Jet 2}. 
{\it Jet 2} shares similar physical characteristics with {\it Jet
1}: (1) a sinuous trajectory, (2) a similar physical width and
length, (3) a similar linewidth of the line core, and (4) a point reflection
symmetry that is especially clear at the center of the system (Figure
\ref{mom0}).  
However, the radial velocity of {\it Jet 2} is
closer to the systemic velocity than for {\it Jet 1}. Another difference is that 
{\it Jet 1} shows in general, lines with more extended wings than {\it Jet 2}.

In Section \ref{sec:3dmodelfit}, we present a numerical fit to the
L1157 CO outflow based on a two precessing molecular jet model, which is in contrast to the wide-angle wind scenario. We
propose this modeling as a proof of feasibility of a multiple jet
model, although this kind of system has been recently proposed for
another Class 0 object, NGC 1333 IRAS 4A2 \citep{Soker:2013zr}. In
that case, the authors tried to explain the observed velocity
gradient in the gas of the outflow \citep{Choi:2011ly}, which is
perpendicular to the outflow axis as produced by the ejection of
multiple jets.

In the two molecular jet scenario, the jets can be launched at the
same time or at different epochs.  A system simultaneously ejecting
two jets may be possible if there are two protostars ejecting two
corresponding jets \citep{Murphy:2008tw}. A striking example of a
binary system (or triple) in which both protostars are launching
bipolar precessing ionized jets is L1551 IRS 5
\citep{Rodriguez:2003qa,Pyo:2009mi,Lim:2006zt,Itoh:2000jl,Fridlund:1998il}.
This system has a separation of 45~AU between the two circumstellar
disks, each of 10~AU in radius. Recent VLA observations of L1157
\citep{Tobin:2013gb} have shown a compact although somewhat elongated
disk-like structure with a radius of $\sim$15~AU, indicating that
only one protostar is present in the system at a 12~AU resolution.
On the other hand, a re-evaluation of the same data with a weighting
more sensitive to find structures (i.e., robust = 0) suggests that
there could be a binary system with $\sim 15$ AU separation (J. Tobin, private
communication).  At this time, we cannot determine if L1157 is a
close binary system or not.

In the case that L1157 is a single protostellar system, the two molecular
jets could be ejected at different times, which could explain why
the CO emission of {\it Jet 2} is weaker than that of {\it Jet 1}
and why {\it Jet 2} does not show SiO emission (thought to disappear
in older and slower shocks, see e.g., \citealt{1999Codella,2006Miettinen,2013FernandezLopez}). 
Thus, {\it Jet 2} would be a remnant
of an older ejection that entrained the molecular environment.  Although L1551 IRS 5 observations support the scenario of two jets, there is one noticeable difference between the system and L1157. The
former has jets with a markedly dissimilar radial velocity (separated
by $\sim$140\kms), possibly indicating two well separated ejection
axes, while in L1157 the two jets show similar radial velocities, which may indicate 
similar ejection axes with only a small inclination difference.
This axis would be 
inclined differently which may be due to secular motions of the system itself.  
On the other hand, there could be different ejection mechanisms launching
multiple jets simultaneously, e.g., a circumstellar disk with an
accreting spiral pattern of multiple arms launching a jet each
\citep[see][]{Soker:2013zr}. This discussion lies beyond
the scope of this paper.

\subsection{3D model fitting to the CO
emission}\label{sec:3dmodelfit}

As stated in the previous section, we suggest that L1157 may be
ejecting two precessing bipolar jets, which we called {\it Jet
1} and {\it Jet 2}. For each jet, we fit the data using a precessing
jet model put forward in \citet{Raga:2009mb}.  This model consists
of a jet launched by a disk-protostar system with a precession
determined by half the opening angle ($\alpha$) and a characteristic
frequency, $\omega$. The bipolar jet is affected by the precession
of the launching system and ejects material running into ballistic
trajectories with constant velocity, $v_j$. For this analysis, we
did not consider additional perturbations due to an orbital binary
motion since we do not know if L1157 is a binary system and this
kind of perturbation displays a pattern only discernible at smaller
spatial scales \citep{Raga:2009mb}. In our model, the ejecta
shows a helix-like geometry in which the
radius of the helix gets wider with time. The following equations,
as derived from \citet{Raga:2009mb}, describe the outflow motion
in a reference frame centered in the disk-protostar system \citep[see
Figure 2 in][]{Raga:2009mb}:
\begin{equation}
\left( \begin{array}{c}
         x\\
         y\\
         z\end{array} \right) = (t-\tau)
   \left( \begin{array}{c}
         v_x\\
         v_y\\
         v_z\end{array} \right),
\end{equation}
where the velocity of the ejection is
\begin{equation}
\left( \begin{array}{c}
         v_x\\
         v_y\\
         v_z\end{array} \right) = v_j
        \left( \begin{array}{c}
         \sin{\alpha}\cos{(s_{prec}\omega\tau-\phi_0)}\\
         -\sin{\alpha}\sin{(s_{prec}\omega\tau-\phi_0)}\\
         \cos{\alpha}\end{array} \right).
\label{eq_velxyz}
\end{equation}

In these equations, $t$ is the present time and $\tau$ is the \lq\lq
negative\rq\rq\ time at which a parcel of the jet was launched such
that $t=0$ means \lq\lq now\rq\rq. $s_{prec}$ defines the direction
of rotation
with +1 indicating clockwise (when seen from top of the Dec axis) and --1 indicating
counter-clockwise rotation. $\phi_0$ is the phase angle in the
xy-plane of the current ejection at $\tau=0$. Also we allow a possible inclination angle $i$ with respect to the
sky plane and a position angle, $pa$, defined counter-clockwise from
north. Thus, we doubly rotate the \textit{jet system} in
$xyz$ coordinates to get the \textit{plane of the sky} $x''y''z''$ 
coordinates, where $x''$ is in the RA direction pointing west, 
$z''$ is in the Dec direction toward north and $y''$ is 
perpendicular to the plane of the sky, pointing away from us.
Given the data provided by the
observations ($x'', z''$, and $v_y''$) and assuming the velocity 
of the ejection does not change, one can obtain the 
ejection time $\tau$ for each data-point
using:
\begin{equation}
\tau=-\sqrt{\frac{(x'')^2+(z'')^2}{v_j^2-(v_y'')^2}}.
\end{equation}
As can be seen from the parametric equations above, 
$\tau$ can be used as the \textit{parameter}, to derive 
a synthetic model, once a set of free variables ($s_{prec}$, 
$v_j$ independent for each jet lobe, $\alpha$, $\omega$, $i$, $pa$, and $\phi_0$) is 
determined. This can be used to speed-up the fitting process.

\begin{figure}
\begin{center}
\epsscale{0.8}
\includegraphics[angle=0,scale=.50]{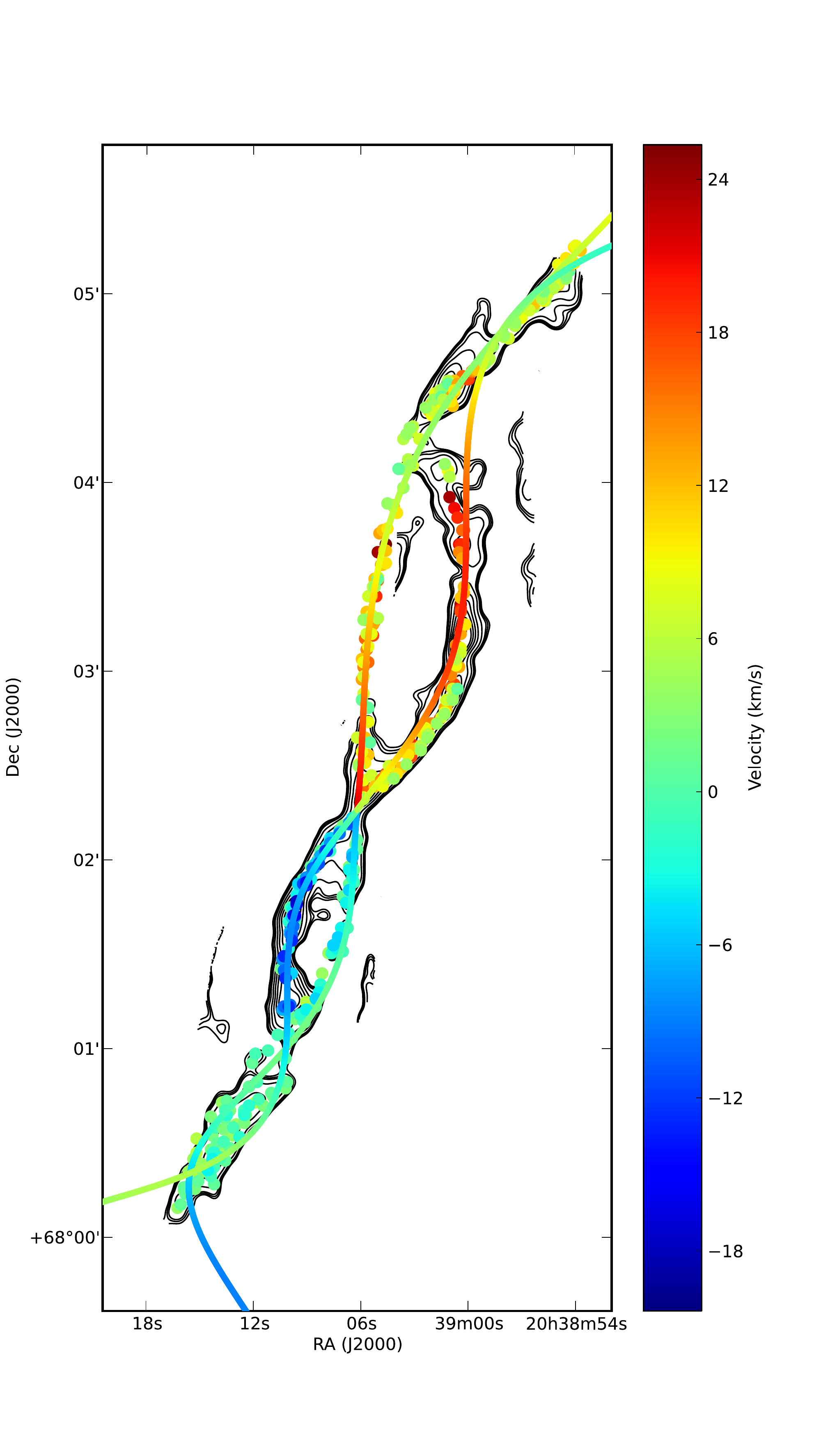}
\caption{Two jet model fit (colored lines) on top of the jet points
used in the fits (colored circles) and the contours representing
the CARMA CO  moment 0 image. The colors of the lines and
points represent the velocity with respect to the cloud velocity.
\label{fit3}}
\end{center}
\end{figure}

\begin{figure}[h]
\epsscale{1}
\plotone{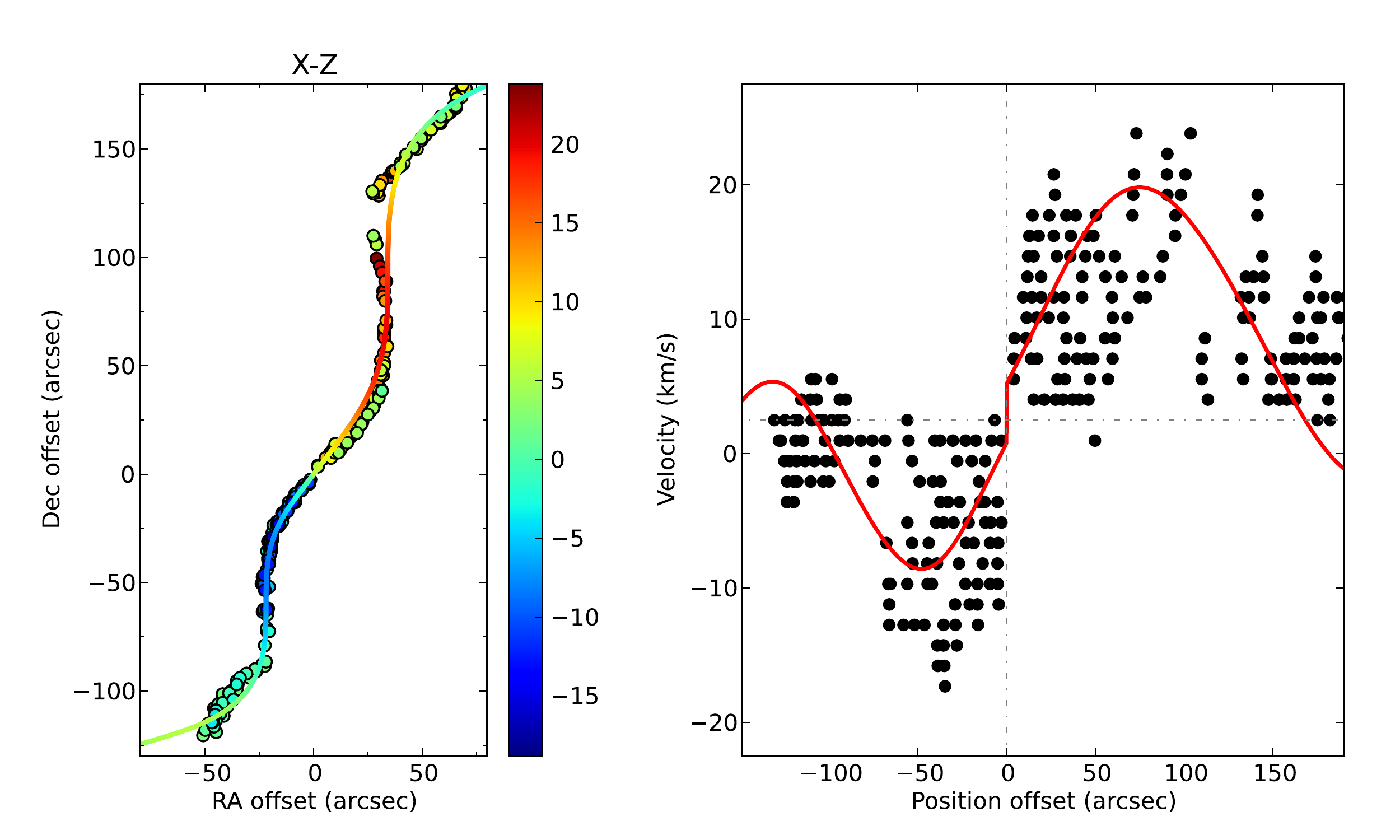}
\caption{{\bf Left:} {\it Jet 1} fit (colored line) on top of the jet
points used in the fit (colored points). The colors represent the
velocity with respect to the cloud velocity. {\bf Right:} The plot
shows the distance from the ejecting source position against the
velocity of each point used to make the {\it Jet 1} fit (black dots). The
red line shows the best fit found for this jet.}
\label{fit1}
\end{figure}

\begin{figure}[h]
\epsscale{1}
\plotone{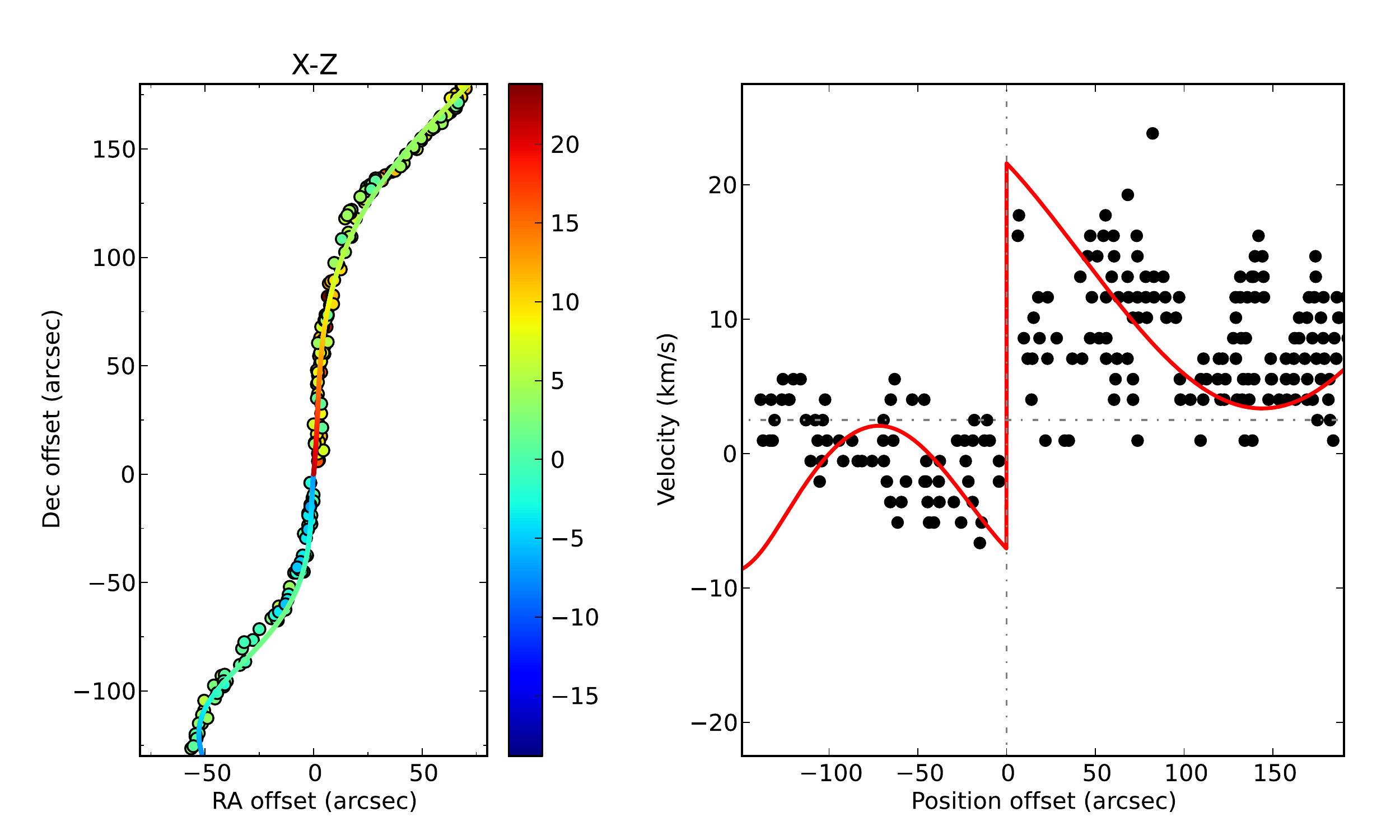}
\caption{The caption is the same as Figure \ref{fit1}, except now
is for the fit of {\it Jet 2}.}
\label{fit2}
\end{figure}

With the model at hand, we extracted the information on position
($x'',z''$) and radial velocity ($v_y''$) of the molecular emission for
each jet by picking out points of peak emission from each 2.5\kms
binned velocity channel image. All the selected points were above
3$\sigma$ and followed an arbitrarily designed jet path.  After
repeating the data selection process several times for each jet,
the final {\it Jet 1} and {\it Jet 2} data points were defined
(Figure \ref{fit3}).  Figures \ref{fit3}, \ref{fit1}, and \ref{fit2}
show the radial velocity and trajectories of the model compared
with the observed data. As can be seen from these figures, there
is good position and kinematic agreement between model and data.
Table \ref{tablefit} shows the values obtained in fitting the two
jets. The model gives higher velocities for the northern lobes
(60--90\kms) than for the southern lobes (30--40\kms) in both jets,
which may be expected since the northern lobe extends farther than
the southern lobe. This could imply, for instance, that different
lobes are ejected at different intrinsic velocities or with different
inclination angles. We choose our model to have two independent 
velocities for each north-going jet and south-going counter-jet. The CO may be tracing the jet 
directly ejected from the surroundings of the central star, or material 
swept up by the jet just in the surroundings of the bullet path, making 
possible differences between the intrinsic velocity in the jet and 
counter-jet. On the other hand, the CO emission may be tracing the 
material swept up by the jet far from the bullet path. In this case, 
\citet{2001A&A...372..899B} proposed that the environment 
surrounding the southern lobe is
denser than that surrounding the northern lobe, thereby reducing
more efficiently the velocity of the southern lobe.
Our model shows
also an agreement between the precessing period of the jets 
(derived from the model, 
they are about $T=$5000-8000 years) and their precessing 
angles ($11\degr$). However, the 
position and inclination angles are different for both jets, 
suggesting a slightly different ejection axis orientation.
The fact that the period and precessing 
angle of both {\it Jet 1} and {\it Jet 2} are similar
could support the scenario of the two jets ejected at 
different times described in Section \ref{tjets}. In such a 
case, one problem is to explain the different ejection axis 
orientation that we previously explained as secular motions 
of the system itself. Such a problem may simply be explained by two protostars each ejecting an outflow.

A note of caution should be made here in establishing that our
fitting is aimed uniquely to probe the feasibility of a two precessing
bipolar jet model in order to explain the observed L1157 CO
emission. We have not tried to univocally fit the model to the data
in any case, and we are aware that other models with multiple jets
could fit the data as well. For instance, {\it Jet 2} has a solution with $s_{prec}=-1$ as good as the one with $s_{prec}=1$ shown here. This is probably due to the poor sampling of {\it Jet 2} in the north lobe and also because the relative weakness of its CO emission compared to that of {\it Jet 1}. We choose not to show the solution with $s_{prec}=-1$ here for consistency on the direction of precession with {\it Jet 1} which has $s_{prec}$ very well constrained. 
Again, even though the model can be
regarded as somewhat arbitrary, its purpose is to qualitatively
show the possibility that the system is composed by at least two
precessing molecular jets and to provide an order of magnitude estimates
of the physical parameters.

\section{Discussion} \label{sec_discussion}\subsection{Possible causes of precession in L1157}\label{subsec_causes}

The cause of the wiggling (precession) pattern of the protostellar jets is
not well known. There are theoretical models explaining this
phenomenon as caused by (1) the orbital motion of a binary system
\citep[e.g.,][]{Masciadri:2002cq}, (2) the precession due to the
tidal interaction between the disk of one protostar and a companion
protostar in a non-coplanar orbit
\citep[e.g.,][]{Terquem:1999kh,Montgomery:2009rq},
(3) the warp of the inner disk (in principle caused by a perturbing
companion star) from which the jet is thought to be launched, and/or
(4) the misalignment between the disk rotation axis and the ejection
engine, with the latter usually understood as an MHD disk-wind
\citep{2014arXiv1402.3553F}.
At this moment there
is no conclusive empirical evidence supporting any of these scenarios. The
first three models would imply a multiple (at least binary)
protostellar system, while model (4) would just require a single
protostar. Nonetheless, in many (if not all) of the protostellar systems
with observed precessing jets, evidence indicate that they
are mostly binary systems (e.g., HH30, \citealt{2008Guilloteau};
HH211, \citealt{2010Lee}; H111, \citealt{1999Reipurth}), and hence 
we do not explore the model of the jet launching engine misaligned
with the disk rotation axis, although we note that it should remain as an
alternative mechanism.
We rule out the orbital motion model
because it entails a mirror symmetry for the jet, a small outflow opening
angle, and a very short period spiraling outflow
\citep{Raga:2009mb}, all of which are
not observed in our data. We cannot test the possibility of a warped inner disk since
we do not have observational data of the dust emission at AU scales. We thus test the
tidal precession model with the L1157 jets. The tidal
precession model has a good theoretical background that we use to
derive the orbital parameters of a hypothetical binary system
producing the precession of the jets.  In particular, we use an
equivalent form of equation (37) from \citet{Montgomery:2009rq} for
circular precessing Keplerian disks. This equation relates the
angular velocity at the disk edge ($\omega_d$), the Keplerian orbital
angular velocity of the companion around the primary protostar
($\omega_o$), and the retrograde precession rate of the disk and the
jet ($\omega_p$):
\begin{equation}
\omega_p=-\frac{15}{32}\frac{\omega_o^2}{\omega_d}\cos{\alpha}\quad,
\end{equation}
where $\alpha$ is the inclination of the orbit of the companion
with respect to the plane of the disk (or obliquity angle), and
this angle is the same as the half--opening angle in equation \ref{eq_velxyz} \citep{Terquem:1999fc}.  The equation also assumes that both objects of the binary
system have the same mass. Introducing the values
for the jet precessing angle $\alpha$ and the precession period
$\tau_p$ derived from our fit (Table \ref{tablefit}) and adopting
some reasonable value for the radius of the disk $r_d$ responsible
for launching the jet, we can constrain  the orbital period $\tau_o$
and the orbital radius $r_o$ of the putative binary system.
\citet{Tobin:2013gb} showed that the disk radius is less than 15~AU,
and we take 1~AU as a lower limit for it. From this we derived
orbital periods and radii in terms of the primary and secondary
protostar's solar masses, $M_1$ and $M_2$: here $M_1=M_2$. For {\it Jet 1} the
orbital period is (50--370)~$M_1^{-1/4}$~yr,
while for {\it Jet 2} the orbital period is 
(60--450)~$M_1^{-1/4}$~yr. The orbital radius for 
{\it Jet 1} is 
(13--52)~$M_1^{1/3}$~AU
 and for {\it Jet 2} the orbital radius is
(15--60)~$M_1^{1/3}$~AU.
It is evident that there is a good agreement between the $\tau_o$
and $r_o$ estimates for {\it Jet 1} and {\it Jet 2}, as expected from
the similar precession periods and angles, and intrinsic velocities. Moreover, the
derived orbital radius is consistent with being a few times the
disk radius which is expected for disks affected by tidal truncation
\citep[e.g.,][]{Artymowicz:1994ss}.  Indeed, using the derived mass
of 0.04\msun for the protostars (Section \ref{sec_rotation}, $M_1 = M_2 = 0.02$\msun), we
then obtain $\tau_o=$130--1200~yr and $r_o=$7--31~AU for a disk with
$r_d=$1--15~AU.

\subsection{Understanding the bipolar outflow kinematics with the envelope}

The precession direction that our models prefer is clockwise when
looking down from north ($s_{prec} = +1$, Table \ref{tablefit}).
Since precession due to tidal interaction of a companion in a non-coplanar orbit
occurs in the opposite direction to the rotation
of the disk (the so-called retrograde precession), the outflow
launching structure would rotate counter-clockwise.  However, our
\ceo and \csto data show a clockwise rotation of the inner envelope
(eastern side redshifted). This implies the disk and the inner
envelope would be counter-rotating.  We discuss how to understand
this counter-rotation of the inner envelope and the outflow launching
structure and the caveats.  When discussing the launching structure,
we specifically mean protostars of a binary system or 
a circumbinary/circumstellar disk.

The precession direction of {\it Jet 1} is significantly well constrained.
In contrast, {\it Jet 2} direction is relatively unclear: the clockwise
and counter-clockwise cases are comparable in our modeling.  This
is probably due to the fact that {\it Jet 2} is not as strong as {\it Jet 1}. 
It is unrealistic to assume that the two jets are precessing in
opposite directions. Therefore, we did not consider this case.  On
the other hand, the inner envelope rotation features discussed with
our \ceo and \csto data are also not decisive due to lack of
angular resolution: the velocity gradient could be contaminated by
the bipolar outflow.  The possibility is low, however, as the
profiles are along the elongated direction nearly perpendicular to
the bipolar outflow, and the same directional velocity gradient in
the inner envelope was also reported by \citet{1997A&A...323..943G}.
Based on the argument above, the counter-rotation of the inner
envelope and the jet launching structure could be accepted.

Such counter-rotation does not seem to be realistic in terms of angular
momentum in star forming molecular core scales. However, some planets
and moons show counter-rotating motions in the solar system and counter-rotating
binary systems may be theoretically feasible, for instance, as a result of the interactions of a
triple system \citep{2014Li}.  In addition, the kinematics of the
L1157 flattened envelope are complicated.  The large-scale solid
body rotation detected in \nthp by \citet{2010ApJ...709..470C} is
counter-clockwise, while their intensity weighted velocity map shows
an overall direction switch at a smaller scale \citep[Figure 3
in][]{2010ApJ...709..470C}.  Counter-rotation indeed helps a
circumbinary disk last longer and is more effective during the binary system
eccentricity evolution \citep{2014arXiv1409.3842D}.  It is not known yet, but L1157 might be a close binary system at 15 AU scale (J. Tobin, private communication).
Although it is not ideal in the
viewpoints of angular momentum, the binary protostellar system and
the circumbinary disk could be counter-rotating. However, the
detailed studies on how such a counter-rotating system forms is
beyond the scope of this paper.

On the other hand, the modeling formalism we used is identical to a rotation
case, so it is possible to interpret the modeling results as a
clockwise rotation instead of precession.  However,  the periods
of several thousand years that our models constrain are too large
for a reasonable launching radius.  Assuming Keplerian rotation
around the protostar, the periods give a 70--80 AU launching radius.

Finally, the jet modeling formalism we used could also be applied to the precession of a misaligned launching structure with respect to the system rotation axis (like the mentioned case 4 on previous section \ref{subsec_causes}), 
so it is possible to interpret the modeling results as a
prograde rather than a retrograde precession.
A prograde precession is also expected in the case of a warped inner disk.
Higher angular resolution and better sensitivity observations will
provide decisive kinematic results toward this interesting Class 0
protostellar system whose bipolar outflow launching structure appears
to be rotating in the opposite direction of the inner envelope.

\section{Conclusions} \label{sec:conclusions}

We present CARMA millimeter observations of the youngest Class 0 protostellar system L1157. The central envelope region (L1157-mm) was imaged in $\lambda=1.3$ and 3~mm continua, CO, C$^{17}$O, CS, CN, $^{13}$CO, C$^{18}$O, and N$_2$H$^+$ with $2\arcsec$ resolution. We also observed CH$_3$OH, H$_2$O, SO, and SO$_2$, but these lines were undetected with the sensitivity of these observations. The continua and the various line features allow us to estimate the physical properties of the envelope. 
In addition, we obtained a large ($5\arcmin$) mosaic image covering the bipolar outflow of L1157 in CO \jto with $5\arcsec$ resolution. 
Our results are summarized in the following:

1. The envelope mass based on the dust continuum flux is estimated
to be $\sim0.583\pm0.022~M_\sun$. We also found that the dust opacity spectral index $\beta$ changes along radius in the image, which is also shown in the visibility data.

2. Among CO isotopes, \tco traces
the CO peaks of the bipolar outflow, while \csto traces the edges.
\ceo shows a structure elongated and perpendicular to the bipolar
outflow on the scale of 100 AU (i.e., it may be tracing the inner envelope).
We also detected a velocity gradient in \csto and \ceo,
which is consistent with Keplerian rotation around a protostar with a mass of
$\sim0.04~M_\sun$ (approximately 1/14 of the envelope mass).
The envelope is rotating clockwise when looking down from the north.

3. \nthp presents double peaks in east-west along the known flattened
envelope direction with a hole at the center.  Each of the double
peaks shows a line profile understood by two velocity components,
which suggests infall motion in the outer envelope of 1000 AU scales.

4. From the bipolar outflow mosaic image in CO,
we detected several ballistic ejections in the redshifted (northern) outflow that support multiple jets. We find that the bipolar outflow could be interpreted as two jets for a reliable example, and we constrain this scenario by modeling the data cube. The idea of two jets is supported by the following observational evidence.

i. The morphology of the CO emission agrees with the two jet scenario.
High-angular resolution CO data show two apparent collimated molecular jets in an X-shape distribution close to the protostar.
In addition, the curved features at large-scale are well reproduced by
the precessing two jet model.

ii. The CO brightness distribution is not well explained by single outflow cavity models but can be explained by the two jet scenario.
The two jet scenario can explain the brightness asymmetry between the NW-SE ridge and the NE-SW ridge. The two jet scenario also can explain the large brightness contrast between the edges and inner part of the lobes.

iii. The radial velocity of the NW-SE jet gas is $2-3$\kms higher than the NE-SW jet gas, which is unexpected for a single jet.
Our model of two precessing jets can explain the different radial velocities, as well as positions, of the NW-SE and NE-SW jets.

iv. The spatial distribution of the SiO emission, which has been observed in the NW-SE jet but not in the NE-SW jet \citep{2001A&A...372..899B}, can be explained if these two jets have different ages.

\acknowledgments
We are grateful to CARMA staff and observers for their dedicated work and anonymous referee for valuable comments that allowed us to improve the paper significantly.
Support for CARMA construction was derived from the states
of Illinois, California, and Maryland, the James S. McDonnell
Foundation, the Gordon and Betty Moore Foundation, the Kenneth T.
and Eileen L. Norris Foundation, the University of Chicago, the
Associates of the California Institute of Technology, and the
National Science Foundation (NSF).
Ongoing CARMA development and operations
are supported by NSF under a cooperative
agreement, and by the CARMA partner universities.
L.W.L. acknowledges NSF AST-1139950.

Facilities: \facility{CARMA}, \facility{SST(IRAC)}.

\bibliographystyle{apj}
\bibliography{L1157_outflow}

\begin{deluxetable}{lcccl}
\tabletypesize{\small}
\rotate
\tablecaption{CARMA observations toward L1157 \label{tab_obs}}
\tablewidth{0pt}
\tablehead{
\colhead{Dates (UT)} & \colhead{Config.} &
\colhead{Calibrators (flux/gain)} &
\colhead{Pointings} & \colhead{Species}
}
\startdata
2010 Jul 22 & E-array &  MWC349, 1927+739 &
outflow (mosaic) & 1 mm Cont., CO, \csto, CN, \ethanol, \water \\
2010 Jul 23 & E-array &  MWC349, 1927+739 &
outflow (mosaic)  & 1 mm Cont., CO, \csto, CN, \ethanol, \water \\
2010 Aug 04 & D-array &  MWC349, 1927+739 &
envelope & 1 mm Cont., CO, \csto, CN, \ethanol, \water \\
2010 Aug 05 & D-array &  Uranus, 1927+739 &
envelope & 1 mm Cont., CO, \csto, CN, \ethanol, \water \\
2010 Oct 17 & C-array &  MWC349, 1927+739 &
envelope & 3 mm Cont., (\ceo, \tco, \nthp, \ethanol, SO, \sot) \\
2010 Oct 19 & C-array &  MWC349, 1927+739 &
envelope & 3 mm Cont., \ceo, \tco, \nthp, \ethanol, SO, \sot \\
2010 Oct 25 & C-array &  MWC349, 1927+739 &
envelope & 3 mm Cont., \ceo, \tco, \nthp, \ethanol, SO, \sot \\
2011 Jul 26\tablenotemark{a} & E-array &  MWC349, 1927+739 &
envelope & 1 mm Cont., CO, CS
\enddata
\tablenotetext{a}{Polarimetric data presented in \citet{2013ApJ...769L..15S}}
\end{deluxetable}

\clearpage
\begin{deluxetable}{lccccl}
\tabletypesize{\small}
\rotate
\tablecaption{Observational results of molecular lines toward L1157
\label{tab_detect}}
\tablewidth{0pt}
\tablehead{
\colhead{Molecular lines} &
\colhead{Frequencies} &
\colhead{Beams} &
\colhead{Velocity Res.} &
\colhead{RMS} &
\colhead{Notes}\\
 & \colhead{GHz} & \colhead{$''\times''$ (PA$\degr$)} &
 \colhead{km s$^{-1}$} & \colhead{Jy bm$^{-1}$ (ch$^{-1}$)}
}
\startdata
1mm continuum  & 230.1\tablenotemark{1} & $2.8\times2.1$ (-18) &   & 0.0047 \\
CO J=2--1   & 230.538000  & $4.8\times4.2$ (72) & 0.51 & 0.34
                            & data mapping the outflow\\
CO J=2--1   & 230.538000  & $2.5\times2.0$ (-17) & 0.50 & 0.10
                            & data toward the envelope\\
\csto\ J=2--1 & 224.714368 & $2.6\times2.2$ (-15) & 0.52 & 0.091
                            & detected in the envelope\\
CS J=5--4    & 244.93556 & $6.3\times2.6$ (-28)   & 0.40 & 0.31
                            & detected northwest of envelope \\
CN N,J,F=2,3/2,5/2--1,1/2,3/2 &  226.659543 & $2.6\times2.1$ (-17) & 0.52 & 0.098
                            & detected in the envelope\\
\ethanol\ 8(-1,8)--7(0,7)     & 229.75876   & $2.5\times2.0$ (-15) & 0.51 & 0.12
                            & not detected \\
\water\ $v_2$=1 5(5,0)--6(4,3)  & 232.68670   & $2.5\times2.1$ (-16) & 0.50 & 0.10
                            & not detected \\
3mm continuum  & 101.4\tablenotemark{1} & $2.1\times1.8$ (-74) &   & 0.00044 \\
\nthp\ 1-0 F$_1$=2--1 F=2-1 \tablenotemark{a} & 93.173480   & $2.3\times2.0$ (-75) & 0.1 (0.065)\tablenotemark{2} & 0.083
                            & detected in the envelope\\
\tco\ J=1--0 & 110.201354  & $1.9\times1.7$ (-74) & 0.22 & 0.079
                            & detected in the envelope\\
\ceo\ J=1--0 & 109.782176  & $1.9\times1.7$ (-75) & 0.1 (0.056)\tablenotemark{2} & 0.084
                            & detected in the envelope\\
\ethanol\ 3(1,3)--4(0,4) A++  & 107.013770  & $2.0\times1.7$ (-76) & 0.23 & 0.075
                            & not detected \\
SO N,J=3,2--2,1   & 109.252212  & $2.0\times1.7$ (-76) & 0.22 & 0.081
                            & not detected \\
\sot\ 3(1,3)--2(0,2) & 104.029410  & $2.0\times1.7$ (-76) & 0.23 & 0.065
                            & not detected \\
\enddata
\tablenotetext{a}{Note that all the 7 hyperfine structure lines
were observed in a band and this line is used as the reference
frequency for the velocity.}
\tablenotetext{1}{The continuum frequencies are a representative value of all continuum data taken by multiple wide spectral windows.}
\tablenotetext{2}{The values in parentheses are velocity resolutions of the original data, which have been re-binned by 0.1 \kms for studies in this paper.}
\end{deluxetable}

\begin{deluxetable}{lcc}
\tablewidth{0pc}
\tablecolumns{3}
\tablecaption{Model fitting parameters
\label{tablefit}}
\tablehead{
\colhead{} & \colhead{{\it Jet 1}} & \colhead{{\it Jet 2}}
}
\startdata
$s_{prec}$             & +1 & +1 \\
$v_{j}$~(\kms)\tablenotemark{a}   & --38$\pm$10 / 59$\pm$12 & --32$\pm$12 / 85$\pm$24 \\
$\alpha$ (\degr)  & 11$\pm$2 & 11$\pm$3 \\
$T$ (yr)\tablenotemark{b}        & 5200$\pm$1300 & 7600$\pm$3500 \\
i (\degr)             & 6$\pm$1 & 11$\pm$2 \\
pa (\degr)            & 26$\pm$2 & 14$\pm$4 \\
$\phi_0$ (\degr)  & 340$\pm$5 & 138$\pm$30 \\
\enddata
\tablenotetext{a}{Jet velocities for the southern (negative) and
northern (positive) lobes}
\tablenotetext{b}{{Large} uncertainties are expected for the precessing
period $T$ and the jet velocity since these parameters cannot be
constrained separately by the model \citep[see][]{Raga:2009mb}.}
\end{deluxetable}

\clearpage

\end{document}